\documentclass[letterpaper,english]{achemso}
\usepackage[T1]{fontenc}
\usepackage[latin9]{inputenc}
\usepackage{xcolor}
\usepackage{amsmath}
\usepackage{amssymb}
\usepackage{graphicx}
\PassOptionsToPackage{normalem}{ulem}
\usepackage{ulem}

\makeatletter

\title{An Analysis of the Gel Point of Polymer Model Networks by Computer
Simulations}
\author{M. Lang, T. Müller}
\affiliation{Leibniz Institut für Polymerforschung Dresden, Hohe Straße 6, 01069
Dresden, Germany}
\email{lang@ipfdd.de}

\providecommand{\tabularnewline}{\\}
\providecolor{lyxadded}{rgb}{0,0,1}
\providecolor{lyxdeleted}{rgb}{1,0,0}
\DeclareRobustCommand{\lyxsout}[1]{\ifx\\#1\else\sout{#1}\fi}

\setkeys{acs}{articletitle = true}

\makeatother

\usepackage{babel}
\begin{document}
\begin{abstract}
The gel point of end-linked model networks is determined from computer
simulation data. It is shown that the difference between the true
gel point conversion, $p_{\text{c}}$, and the ideal mean field prediction
for the gel point, $p_{\text{c,id}}$, is a function of the average
number of cross-links per pervaded volume of a network strand, $P$,
and thus, contains an explicit dependence on junction functionality
$f$. On the contrary, the amount of intra-molecular reactions at
the gel point is independent of $f$ in a first approximation and
exhibits a different power law dependence on the overlap number of
elastic strands as compared to the gel point delay $p_{\text{c}}-p_{\text{c,id}}$.
Therefore, $p_{\text{c}}-p_{\text{c,id}}$ cannot be predicted from
intra-molecular reactions and vice versa in contrast to a long standing
proposal in literature. Instead, the main contribution to $p_{\text{c}}-p_{\text{c,id}}$
for $P>1$ arises from the extra bonds (XB) needed to bridge the gaps
between giant molecules separated in space and scales roughly $\propto\left(P-1\right)^{-1/2}$.
Further corrections to scaling are due to non-ideal reaction kinetics,
composition fluctuations, and incompletely screened excluded volume,
which are discussed briefly.
\end{abstract}

\section*{Introduction}

Polymer networks and gels are materials that have reached a wide range
of applications ranging from car tyres to drug delivery, removal of
pollutants, artificial muscles, or stretchable electronics \cite{Samaddar2019,Farhood2019,You2019,Qiu2019}.
One point of major interest for theory, processing and application
is the exact location of the gel point \cite{Flory1953,Korolev1982},
since there, the reacting liquid turns into a solid. The properties
next to the gel point are well understood for critical percolation
and mean field models as a function of the distance to the gel point.
The prediction of the percolation threshold or the gel point itself
remains a challenging problem for theorists \cite{Mark1997,Rubinstein2003,Mertens2018,Torquato2013}.
Typically, the classical Flory-Stockmayer (FS) theory \cite{Flory1941a,Stockmayer1943}
or an equivalent mean field model \cite{Macosko1976,Miller1976} is
taken for a first estimate based upon an ideal system whereby both
intra-molecular reactions prior to gelation and the positions of the
reacting molecules in space are neglected. Previous generalizations
of the FS theory \cite{Harris1955,Kilb1958,Ahmad1980,Korolev1982,Suematsu1998,Sarmoria2001,Suematsu2002,Tanaka2012a,Wang2017}
focus essentially on corrections due to intra-molecular reactions
(``loops''), since the self-contacts of random walks or branched
polymers in semi-dilute solutions are well understood. The impact
of the spatial arrangement of the reacting molecules on the position
of the gel point was essentially ignored in literature due to the
lack of an analytical approach that allows for a quantitative treatment
of this point.

One remarkable result of these models (mean field + loop correction)
is that the latest variant of it \cite{Wang2017} seems to work even
for overlap numbers around one and below, which is in the core of
the critical percolation regime or even requires diffusion of the
molecules to allow for network formation. On the contrary, there is
a number of simulation works \cite{Lee1990,Gupta1991,Dutton1994,Hendrickson1995,Lang2007,Yang2007}
(see section ``Numerical studies in literature'' of the Appendix
for a more detailed discussion) that indicate that the loop correction
might not be sufficient to explain the delay of the gel point for
overlap numbers clearly above one. These works, however, could be
criticized, since network formation was modeled either without diffusion
of the reactive species \cite{Lee1990,Gupta1991,Dutton1994,Hendrickson1995},
or analyzed only indirectly \cite{Lang2007}, or some inconsistency
among the data is apparent \cite{Yang2007}. Nevertheless, the simulation
studies \cite{Lee1990,Gupta1991,Dutton1994,Hendrickson1995,Lang2007,Yang2007}
are in line with the scaling model of percolation, where the hyperscaling
relation connects the fractal dimension of the branched molecules
to space dimension \cite{Rubinstein2003}: extra reactions are necessary
to bridge the gaps between the separated giant molecules just below
the gel point. This is not accounted for in the mean field models.
It should be observable as a gap between the conversion at the true
gel point and a mean field estimate for the gel point were corrections
due to intra-molecular reactions were considered.

Using the jargon of percolation, the gelation of $f$-functional stars
(GS) - or equivalently, the end-linking of $N$-mers through $f$
functional junctions - is a non-nearest neighbor percolation problem
of almost randomly distributed sites with limited ``valence'' (the
junction functionality $f$). Originally, any kind of non-nearest
neighbor percolation problems was termed ``long-range percolation''
\cite{Hoshen1978,Stephen1980,Ord1982,Ray1988,Meester1994,Frei2016}.
This nomenclature has partially changed in recent years, since a distinction
among qualitatively different ``long-range'' models could be made.
The general finding is that an exponential cut-off or a hard range
limit for the bonds maintains a window of critical percolation next
to the critical point. However, the range of conversion where critical
percolation can be observed might be rather narrow, if a large number
or neighbors can be reached \cite{Aharony1982}. Therefore, it was
speculated that the narrow range of conversions with critical properties
may not be accessible experimentally \cite{Ord1982}. This motivated
to estimate the gel point delay by mean field arguments, also, since
it was shown that the size distribution of the smallest loops near
the gel point follows mean field statistics \cite{Lang2005a,Wang2017}.

Nevertheless, GS falls within this class of ``short-range'' problems,
since the end-to-end distribution of the bonds (the polymer strands)
is characterized by an exponential decay. Power law decays for the
bond length distribution can lead to a short range behavior, to a
small-world behavior for a certain window of power laws depending
on space dimension, or to a true long-range behavior, which is mean-field
like on all scales \cite{Sen2002,Luijten2002}. A limited valence
of the nodes has no effect \cite{Gaunt1979,Kertesz1981} on the critical
behavior once $f\ge3$, but affects certainly the location of the
critical point. Thus, the general applicability of percolation to
model GS in the vicinity of the gel point is out of question and has
been corroborated by more recent works on percolation within a range
\cite{Qian2016,Ouyang2018,Deng2019} (some newer works call this type
of problems ``medium-range percolation'' \cite{Deng2019} or ``equivalent
neighbor percolation'' \cite{Qian2016}). 

Nevertheless, there are several points that are different in GS as
compared to the percolation problems that were studied previously:
\begin{enumerate}
\item Reactions between molecules occur based upon a diffusion reaction
mechanism. The applicability of critical percolation approximation
requires that diffusion effects are not important.
\item The position of network junctions couples to the local polymer density
once these become connected to chains. Composition fluctuations arise
for cross-linking at the present of a solvent or if different chemical
species are linked together to form a network (e.g. chains and junctions).
These composition/interaction effects can introduce a second characteristic
length scale that may affect the scaling in the vicinity of the gel
point.
\item The number of accessible junctions depends not only on range (here
given by degree of polymerization, chain stiffness, and solvent quality)
but also on the junction functionality, polymer volume fraction, and
the stoichiometric ratio $r$ of reactive groups on junctions vs.
reactive groups on chains.
\end{enumerate}
Indeed, kinetic gelation is one example where the diffusion reaction
mechanism dominates the behavior and causes a different universality
class as compared to critical percolation \cite{Herrmann1986}. Similarly,
diffusion limited cluster aggregation is a different universality
class \cite{Meaking1983} and would become relevant once the overlap
number of stars drops below one. For sufficiently large overlap numbers
it is expected that diffusion becomes unimportant in the close vicinity
of the gel point, since the behavior of the system is dominated there
by the largest molecules, which diffuse extremely slow because of
the high viscosity next to the gel point. What is not accounted for
in this discussion is that the local dynamics and accessibility of
the nodes depends on the number of attached chains. Thus, bonds are
no more inserted randomly into the system, instead, nodes with a smaller
number of existing connections exhibit a higher rate of bond formation.
We analyze this point in the appendix and correct the mean field estimate
for the critical point accordingly.

The second point above is suppressed to a significant extent by analyzing
network formation in good solvents or melts. Also, the homo-polymerization
of stars instead of a co-polymerization of junctions and chains or
stars of two different types is preferable here. In the best possible
case (homo-polymerization of stars), the length scale at which the
local concentration of the nodes couple to a constant polymer density
is comparable to the size of the network strands and thus, irrelevant
for the behavior in the vicinity of the gel point. Nevertheless, it
is certainly of interest to analyze the impact of statistical fluctuations
in composition in case of end-linking reactions, since end-linking
reactions are the literature standard for gelation studies of model
networks. This point is also of relevance for high conversions, since
composition fluctuations freeze in during the cross-linking reaction
\cite{Lang2019}.

The third point is explained best when comparing the average number
of possible neighbors $P$ to which a node can connect in typical
percolation studies with GS. In bond percolation were bonds are introduced
within a range $R$, the number of neighbors $P$ is typically a constant
number of roughly $P\approx\rho R^{3}$, where $\rho$ is the number
density of nodes. On the other hand, the corresponding overlap number
of junctions for the simplest case of GS homo-polymerization of monodisperse
stars with $N/f$ Kuhn segments per arm is given by
\begin{equation}
P\approx\frac{8\pi\phi R^{3}}{3fNb^{3}}.\label{eq:Pj}
\end{equation}
Here, $R$ is the average extension of the network strands that depends
on the number of Kuhn segments, $N$, the root mean square size of
a Kuhn segment, $b$, and the solvent quality. $\phi$ is the polymer
volume fraction and $b^{3}$ a rough estimate for the occupied volume
of a Kuhn segment. Thus, $P$ is an explicit function of the valence
$f$ in case of polymer gelation in contrast to typical long range
percolation studies, where the number of bonds can be up to $\approx\rho R^{3}$.

The above points and the question whether the gel point can be estimated
by a consideration of intramolecular reactions is addressed with our
publication. We simulate explicitly the dynamics, conformations, and
reactions of the molecules that form the network in space to remove
the limitations of previous simulation studies. We determine intra-molecular
reactions and the position of the gel point. On this basis, we demonstrate
that the delay of the gel point is not controlled by intra-molecular
reactions. We show that the dominating contribution is due to extra
bonds (XB) that are necessary to bridge the gaps between the giant
molecules. We discuss corrections that arise due to a different mobility
of the reacting species, due to composition fluctuations, and due
to incompletely screened excluded volume. Furthermore, we provide
examples how the delay of the gel point could be analyzed with more
detail in the experiment. 

\section*{Computer Simulations}

For our study, we use the Bond Fluctuation Model (BFM) \cite{Carmesin1988,Deutsch1991},
which is a well known lattice based Monte-Carlo method that has been
used frequently to simulate polymer model systems \cite{Rabbel2017,Lang2018b,Mueller2019},
solutions \cite{Lang2012,Klos2016}, membranes \cite{Werner2014},
melts \cite{Kreer2001,Meyer2010}, or networks \cite{Lang2013,Lang2016}.
In this simulation method, chains are represented by a connected set
of small cubes that resemble the monomers of the chain. Monomers are
connected into chains (and later: to network junctions) through a
discrete set of 108 different bond vectors. Monodisperse melts made
of $M$ chains with a degree of polymerization $N=8,...,64$ were
equilibrated by random jumps of the monomers under the constraint
that all bonds remain within the allowed set of bond vectors. All
simulations were run on a lattice of $256^{3}$ lattice sites with
periodic boundaries. A stoichiometric amount of $X=2M/f$ network
junctions are also modeled by small cubes and are added at random
positions to the melt of chains such that approximately $2^{20}$
monomers occupy a volume fraction of $\phi\approx0.5$ of the lattice
sites, see Table \ref{tab:The-critical-conversion} for the simulation
parameters. This volume fraction is standard for the simulation of
dense systems like melts or networks using the BFM \cite{Paul1991a,Wittmer2007}.

After equilibration of the reaction mixture, end-linking reactions
were turned on while chain monomers and junctions perform a stochastic
motion inside the reaction bath. Whenever a free chain end was in
one of the nearest neighbor positions to a junction that was not yet
completely reacted, a bond was introduced between both. The chain
end is then bound to the junction and the number of possible further
reactions of the junction is reduced by one. 100 statistically independent
samples (by initial positions of cross-links and chain conformations)
with the same simulation parameters were created and subsequently
linked into networks in order to improve statistics. Note that our
simulations explicitly model a diffusion collision mechanism in space
to simulate reactions. This is different to the random insertion of
bonds either between spatially correlated neighbors as in percolation
studies or only correlated within a single molecule as in recent mean
field work \cite{Wang2017}.

\section*{Mean field estimates of the gel point}

The ideal reference for the gel point of our end-linked model networks
is \cite{Miller1976} 
\begin{equation}
p_{\text{c,id}}=\left(\frac{1}{f-1}\right)^{1/2}\label{eq:pcid}
\end{equation}
and the delay of the gel point conversion with respect to $p_{\text{c,id}}$
is quantified below as $p_{\text{c}}-p_{\text{c,id}}$. The power
of $1/2$ in the above equation results from the two bonds at both
chain ends that are necessary to link the $f$-functional junctions. 

intra-molecular reactions in the reaction bath are determined by counting
the number of connected components, $C$, the total number of bonds,
$B$, between initial molecules with a total number of $M+X$. Then,
the cycle rank (total number of independent circuits in the graph)
\begin{equation}
\xi=B-(M+X)+C\label{eq:cycle rank}
\end{equation}
 provides the total number of intra-molecular reactions in the reaction
bath\cite{Lang2001b}. It has been argued by several authors \cite{Flory1941a,Harris1955,Kilb1958,Ahmad1980,Korolev1982,Suematsu1998,Sarmoria2001,Suematsu2002,Tanaka2012a,Wang2017}
that the formation of finite loops (intra-molecular reactions) controls
the displacement of the gel point with respect to $p_{\text{c,id}}$,
since each intra-molecular reaction diminishes the number of branching
reactions by one. Thus, we estimate the shift of the gel point due
to intra-molecular reactions, $\Delta p$, with respect to the total
number of possible reactions, $2M$, through 
\begin{equation}
\Delta p=\xi/(2M).\label{eq:Dp}
\end{equation}
 Then, 
\begin{equation}
p_{\text{c}}-\Delta p\left(p_{\text{c}}\right)=p_{\text{c,id}}\label{eq:self-consistent}
\end{equation}
provides a self-consistent mean field estimate for the gel point,
$p_{\text{c}}$, that is corrected by intra-molecular reactions.

Below, we will see that such a self-consistent approach underestimates
cyclization at the true gel point. Therefore, the $\Delta p$ data
in Table \ref{tab:The-critical-conversion} were collected at the
true gel point as determined by the onset of a non-vanishing modulus.
The data for a self-consistent determination of $\Delta p$ are included
in Figure \ref{fig:Collapse1} for completeness and available with
high accuracy through the fit of the corresponding data.

\begin{table}
\begin{tabular}{|c|c|c||c|c||c|c|c||c|c|c|}
\hline 
$f$  & $N$  & $X$  & $p_{\text{c,\ensuremath{\mu}}}$  & $\mu$  & $p_{\text{c,\ensuremath{\gamma}}}$  & $\gamma$  & $C_{-}/C_{+}$  & $p_{\text{c,id}}$  & $\Delta p$  & $p_{\text{d}}$\tabularnewline
\hline 
\hline 
3  & 8  & 80653  & $0.750$  & $1.78\pm0.01$  & $0.7506$  & $1.47\pm0.05$  & $3.0\pm0.8$  & $0.7071$  & $0.0213$  & $0.690$\tabularnewline
\hline 
3  & 16  & 41942  & $0.742$  & $1.81\pm0.01$  & $0.7409$  & $1.45\pm0.06$  & $3.0\pm0.9$  & $0.7071$  & $0.0145$  & $0.695$\tabularnewline
\hline 
3  & 32  & 21398  & $0.732$  & $1.81\pm0.01$  & $0.7344$  & $1.46\pm0.07$  & $3.5\pm1.4$  & $0.7071$  & $0.0100$  & $0.700$\tabularnewline
\hline 
3  & 64  & 10810  & $0.725$  & $1.90\pm0.02$  & $0.7272$  & $1.37\pm0.07$  & $2.7\pm0.8$  & $0.7071$  & $0.0070$  & $0.697$\tabularnewline
\hline 
\hline 
4  & 8  & 61680  & $0.636$  & $1.74\pm0.01$  & $0.6371$  & $1.54\pm0.07$  & $3.0\pm1.2$  & $0.5774$  & $0.0225$  & $0.562$\tabularnewline
\hline 
4  & 16  & 31774  & $0.622$  & $1.80\pm0.01$  & $0.6230$  & $1.53\pm0.11$  & $2.0\pm1.4$  & $0.5774$  & $0.0153$  & $0.568$\tabularnewline
\hline 
4  & 32  & 16130  & $0.611$  & $1.86\pm0.02$  & $0.6135$  & $1.38\pm0.07$  & $2.2\pm0.7$  & $0.5774$  & $0.0106$  & $0.568$\tabularnewline
\hline 
4  & 64  & 8128  & $0.601$  & $1.88\pm0.02$  & $0.6041$  & $1.31\pm0.06$  & $2.0\pm0.5$  & $0.5774$  & $0.0073$  & $0.569$\tabularnewline
\hline 
\hline 
6  & 8  & 41942  & $0.523$  & $1.69\pm0.02$  & $0.5235$  & $1.53\pm0.11$  & $3.0\pm1.5$  & $0.4472$  & $0.0239$  & $0.436$\tabularnewline
\hline 
6  & 16  & 21398  & $0.503$  & $1.89\pm0.02$  & $0.5043$  & $1.43\pm0.09$  & $2.0\pm1.2$  & $0.4472$  & $0.0161$  & $0.442$\tabularnewline
\hline 
6  & 32  & 10810  & $0.492$  & $1.78\pm0.02$  & $0.4916$  & $1.37\pm0.05$  & $1.7\pm0.9$  & $0.4472$  & $0.0112$  & $0.444$\tabularnewline
\hline 
6  & 64  & 5433  & $0.481$  & $1.95\pm0.04$  & $0.4813$  & $1.28\pm0.05$  & $1.8\pm0.4$  & $0.4472$  & $0.0077$  & $0.442$\tabularnewline
\hline 
\end{tabular}\caption{\label{tab:The-critical-conversion}Simulation parameters and gel
point estimates. $f$ is the junction functionality, $N$ the degree
of polymerization of linear chains between junctions, and $X$ is
the total number of junctions per simulation. The gel point is estimated
from simulation data in two different ways a) By an extrapolation
of modulus data, $p_{\text{c,\ensuremath{\mu}}}$, with critical exponent
$\mu$ that is fit to the data, and b) by an analysis of the weight
average degree of polymerization $N_{\text{w}}$ as $p_{\text{c,\ensuremath{\gamma}}}$
with an exponent $\gamma$. $C_{-}/C_{+}$ is the ratio of the coefficients
of the branches of $N_{\text{w}}$ below and above the gel point.
The conversion lost in intra-molecular reactions, $\Delta p$, is
measured at $p_{\text{c,\ensuremath{\mu}}}$. $p_{\text{d}}$ is the
mean field estimate for the gel point that reflects unequal reactivity,
while $p_{\text{c,id}}$ assumes binomially distributed connections
on junctions and chains. The standard deviation for $p_{\text{d}}$
is below $10^{-3}$, the error for $\Delta p$ is below $3\times10^{-4}$
while the error for the determination of $p_{c,\mu}$ is about $10^{-3}$.
For $p_{\text{c},\gamma}$, we require four digits behind the comma
to fix the numerical optimum of a parallel decay of both branches,
but the last digit is certainly not significant as visible from the
large error for $\gamma$ and the scatter of the $N_{\text{w}}$ data.}
\end{table}

Each intra-molecular reaction creates a cyclic structure inside the
network (``loop''). We have analyzed loop size distributions in
section ``intra-molecular reactions'' of the Appendix. Our results
show that the size distribution of loops is $\propto i^{-5/2}$ in
a good approximation next to the true gel point, where $i$ is the
number of precursor chains that establish the loop. In consequence,
we expect that the total amount of loops at the gel point is independent
of $f$ up to minor corrections for smallest $i$, since a dependence
$\propto i^{-5/2}$ refers to a mean field gel point where there is
in average exactly one path to the infinite gel in the limit of large
$i$ (see also equations (20) to (24) of Ref. \cite{Lang2005a} or
equation (A2-44) of Ref. \cite{Dusek1978}). A mean field scaling
for the short range statistics is expected for overlap numbers $\ge1$
and has been found previously in non-nearest neighbor bond percolation
\cite{Ray1988,Hoffmann2011}. 

A rough estimate of the generation $i$ at which a cross-over to critical
percolation is expected can be made through the Ginzburg criterion.
Let us introduce the relative extent of reaction, 
\begin{equation}
\epsilon=\left(p-p_{\text{c}}\right)/p_{\text{c}},\label{eq:epsilon}
\end{equation}
as a measure of conversion $p$ with respect to the conversion at
the true gel point, $p_{\text{c}}$. According to experimental data
\cite{Colby1992} for $f=4$ (same $f$ as in Figure \ref{fig:size-dist}),
the value of $\epsilon$ at the cross-over to critical percolation,
$\epsilon_{\text{G}},$ is $\epsilon_{\text{G}}\eqsim N^{-1/3}/10$,
which provides $\epsilon_{\text{G}}\approx1/20$ for the data of Figure
\ref{fig:size-dist}. We expect mean field scaling up to this average
number of strands $i\approx20$. The amount of loops in generations
$i\ge20$ contributes only a minute portion to the total number of
loops in the range of 1-2\% depending on whether mean field or critical
scaling is assumed for the loop size distribution. Thus, the total
amount of intra-molecular reactions is rather accurately estimated
by adopting a mean field model for the loop size distribution in the
vicinity of the gel point. We further note that our simulation results
for $\Delta p$ represent formally an upper bound for macroscopic
systems, since a minute part of cyclic structures is closed through
the periodic boundary conditions (finite size effect). These artificial
giant loops involve of order $L^{2}/b^{2}\approx10^{4}$ segments
and thus, loops with $i\gtrsim10^{4}/N$ chains, which yields a contribution
to $\Delta p$ below the statistical error. Therefore, finite size
corrections can be ignored for $\Delta p$.

In a more general form, the width of the percolation regime $|p-p_{\text{c}}|/p_{\text{c}}$
for long range percolation can be estimated \cite{Aharony1982} as
a function of the number of possible reaction partners $P$
\begin{equation}
\epsilon_{\text{G}}\cong P^{-2/(6-d)}.\label{eq:Toulouse}
\end{equation}
The width of the critical zone refers to a relative extent of reaction
in the range of (after dropping coefficients of order unity and using
equation (\ref{eq:Pj}))
\begin{equation}
\left|\frac{p-p_{\text{c}}}{p_{\text{c}}}\right|\lesssim\left(\frac{f}{\phi N^{1/2}}\right)^{2/3}.\label{eq:Scaling}
\end{equation}
This leads to the well known result of $\epsilon_{\text{G}}\propto N^{-1/3}$
for melts with $\phi=1$ and $R\approx bN^{1/2}$, but indicates also
that $\epsilon_{\text{G}}\propto f^{2/3}$.

A second possible deviation from mean field results from a different
mobility and accessibility of junctions (and eventually chain ends)
as a function of the number of existing connections to other molecules.
In similar manner, concentration fluctuations of junctions and chain
ends may affect the distribution of connections of the molecules.
In consequence, this connectivity distribution is no more a binomial
one, which we indeed observe in our simulations, see section ``Unequal
reactivity'' of the Appendix. This difference to the ideal case can
be considered within a mean field approach by mapping the system to
a co-polymerization of junctions and chains with functionality distributions
that equal the distribution of the number of connections at a given
conversion $p$ as described in more detail in the Appendix. Let $f_{\text{e}}$
denote the weight average number of connections among the junctions
and $g_{\text{e}}$ the weight average number of connections of chains.
Then, the gel point reflecting the non-ideal distribution of connections
is \cite{Macosko1976} the conversion $p_{\text{d}}$ where $(f_{\text{e}}-1)(g_{\text{e}}-1)=1$
for the stoichiometric systems of our study. This condition is determined
numerically from the simulation data and given in Table \ref{tab:The-critical-conversion}.
The corresponding mean field estimate for the gel point that reflects
both non-ideal reactivity and intra-molecular reactions is given in
good approximation by $p_{\text{d}}+\Delta p$, see Table \ref{tab:The-critical-conversion}.

For all networks of our study, the impact of lower accessibility and
mobility of junctions causes a smaller $f_{\text{e}}$ next to the
gel point as compared to the binomial reference case. On the contrary,
$g_{\text{e}}$ seems to be dominated by concentration fluctuations
as the portion of chains with two connections increases quicker than
$p^{2}$. In total, most of these two competing corrections compensates
each other and causes only a small shift of the estimated gel point
towards lower conversions, $p_{\text{d}}<p_{\text{c,id}}$. Thus,
$p_{\text{c},\text{id}}+\Delta p$ is an upper bound for a mean field
estimate with corrections due to intra-molecular reactions. This observation
holds for all samples of our study and indicates that composition
fluctuations are more relevant than mobility or accessiblity of the
reactive groups, at least for the low $f\le6$ of our study. Furthermore,
we have to point out that our simulations were performed in the diffusion
controlled regime where the impact of mobility is expected to be largest.
Experiments are typically reaction controlled. Thus, a quantitatively
larger shift of $p_{\text{d}}$ towards lower conversion can be expected,
in particular, since our simulations refer also to the limiting case
of a perfect mixture. Based upon this observation, we recommend that
experimental tests of theory should be conducted with a homo-polymerization
of $f$-functional stars as model systems, since this architecture
diminishes the impact of a different mobility as only chain ends react.
Also, concentration fluctuations of reactive groups are suppressed
to a length scale comparable to the size of the stars, if reactions
are conducted in a sufficiently good solvent or in melt. 

\section*{Gel point estimate based upon phantom modulus}

The gel point can be determined from the onset of a non-vanishing
equilibrium modulus of the networks. In the vicinity of the gel point,
the phantom contribution to modulus dominates the elasticity \cite{Rubinstein1994}.
We analyze the phantom modulus as suggested in Ref. \cite{Gusev2019}
by considering that each chain is an ideal elastic spring of stiffness
$\propto N^{-1}$. This direct determination is preferable as compared
to counting the cycle rank, since the elastic contribution of a network
cycle depends on its size and embedding into the network structure
\cite{Zhong2016,Lang2018,Panyukov2019,Lang2019b}: network strands
in small loops (by number of strands) contribute in average less to
modulus than strands in large loops. The classical limit is reached
for loops made of an infinite number of strands (infinite tree approximation).
The origin for this difference is that the extension of a strand of
$N$ segments within a cyclic polymer of $iN$ segments is in average
smaller than in its linear counterpart. Since modulus is the free
energy change with respect to elongation, it is the time average size
of the strands that determines the contribution to modulus (and not
only network connectivity as assumed when considering the cycle rank)
\cite{Lang2018,Lang2019b}. Qualitatively, this behavior is somewhat
similar to the behavior of de-swollen gels \cite{Obukhov1994} where
the reduced chain extension in the dry state causes a significant
reduction of modulus as compared to gels that were prepared in the
dry state.

\begin{figure}
\includegraphics[width=1\columnwidth]{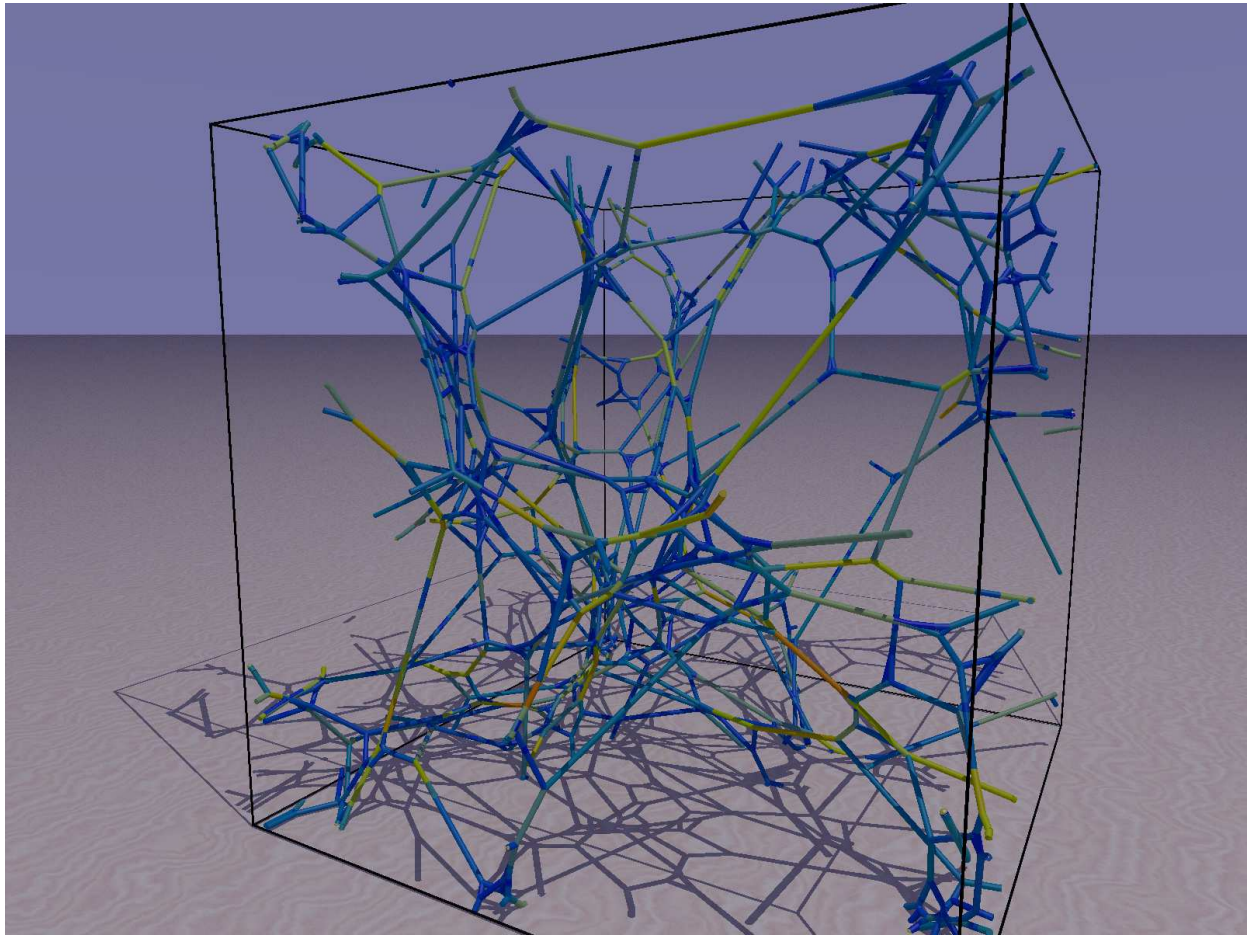}

\caption{\label{fig:The-force-balance}The force balance condition for the
determination of the phantom modulus of a network with $f=4$ and
$N=8$ is shown at $p=0.67$ not far beyond the gel point. The color
code (from blue to red) corresponds to a different strain of the strands.
Note that only the elastically active material is shown and that single
strands between branch points correspond typically to a sequence of
several linear chains or loops, since $p$ is close to the gel point.
A change in color along one straigth connection indicates, therefore,
a change in topology (for instance, a linear strand followed by a
stretched loop where the loop strands carry less load, etc ...). }
\end{figure}

For a network of ideal springs, the ground state is determined numerically
by considering a simultaneous force balance at all junctions. Below,
the gel point, the ground state refers to a collapse of the structure
into a single point. Above the gel point, the periodic boundaries
of the sample prevent this collapse and a non-zero size of the springs
is obtained. An example for the force balance condition to determine
modulus is shown in Figure \ref{fig:The-force-balance} for a network
with parameters $f=4$ and $N=8$ at $p=0.67$. Finally, the resulting
elastic energy density within the samples is averaged over the 100
copies of equivalent networks to determine the average phantom modulus
of the samples.

\begin{figure}
\includegraphics[angle=270,width=1\columnwidth]{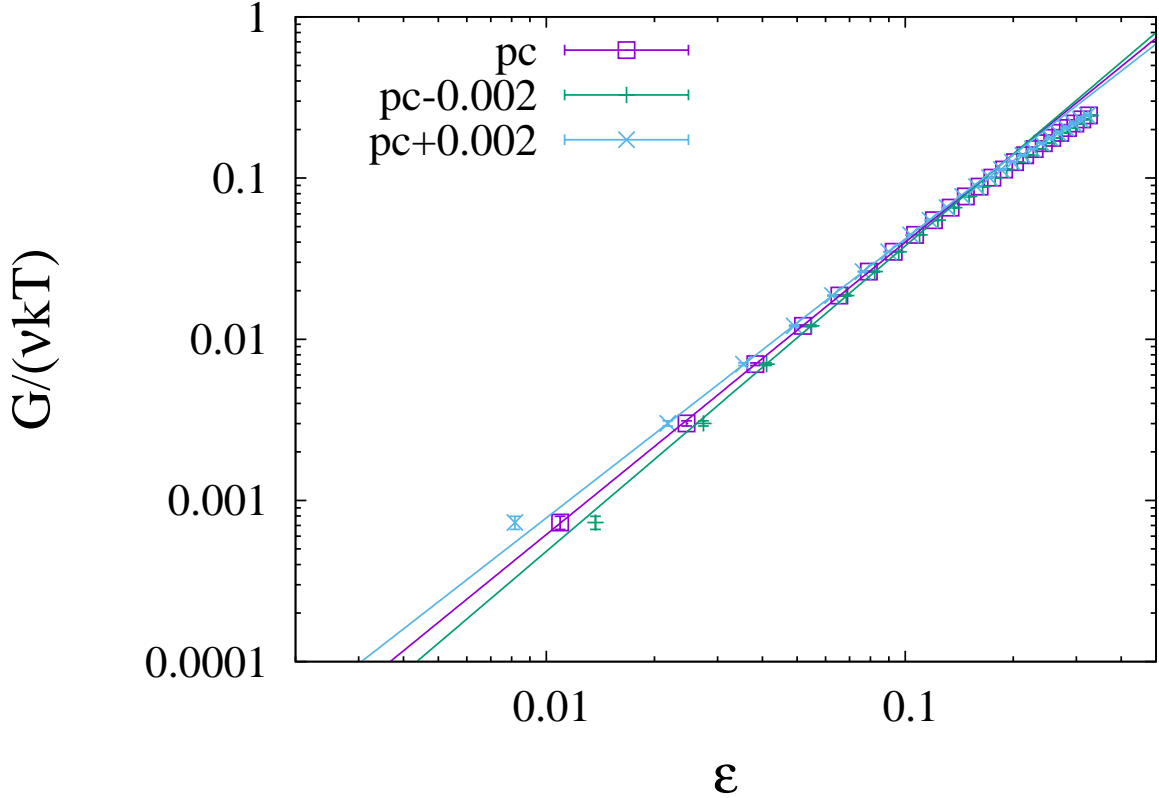}

\caption{\label{fig:Scaling-of-phantom}Scaling of phantom modulus in the vicinity
of the gel point for the networks with $f=3$ and $N=32$. A presumed
$p_{\text{c},\mu}$ that is either 0.002 below or above the optimum
$p_{\text{c}}$ already leads to significant deviations (beyond error
bar) from a power law dependence for the smallest $G$ values.}
\end{figure}

In the scaling model of gelation, it is expected that phantom modulus
grows in the vicinity of the gel point as a power law 
\begin{equation}
G\propto\epsilon^{\mu}.\label{eq:G}
\end{equation}
In order to detect the gel point, we vary $p_{\text{c}}$ in steps
of $10^{-3}$ and fit all data with a modulus in the range $10^{-4}G_{\text{max}}<G<G_{\text{max}}/20$
to a power law $\epsilon^{\mu}$. Here, $G_{\text{max}}$ is the maximum
modulus at maximum conversion. The boundaries for the fit serve to
reduce noise effects at the lower bound and to exclude the cross-over
to mean field at large $G$. The gel point is identified as the conversion
with the lowest deviation to a power law dependence in the above range
with focus on smallest $G$ values and denoted below as $p_{\text{c,\ensuremath{\mu}}}$.
Typically, a clear power law dependence over one and a half decades
is found for the smallest $G$ values for a narrow range of conversions
with a width of approximately $10^{-3}$ around the ``gel point''.
Figure \ref{fig:Scaling-of-phantom} shows a typical example for the
detection of this $p_{\text{c,\ensuremath{\mu}}}$ plus two examples
with a slightly deviating estimate for the gel point. The exponent
$\mu$ is taken as an adjustable parameter and fit to the data. The
results of these fits for $p_{\text{c,\ensuremath{\mu}}}$ and $\mu$
are summarized in Table \ref{tab:The-critical-conversion}.

For mean field, $\mu=3$ is expected \cite{Gordon1969} in agreement
with scaling predictions for 6 dimensions. For critical percolation,
one expects for $d=3$ according to \cite{Skal1975,DeGennes1976,Sahimi1998}
$\mu=\zeta+\nu\left(d-2\right)$ that there is approximately $\mu\approx1.9$.
Estimates for the exponent $\nu$ that describes the divergence of
the correlation length were originally \cite{Isichenko} about $\nu\approx0.9$,
while more recent renormalization group estimates or simulation data
provide slightly lower values of $\nu\approx0.82$ or $\nu\approx0.875$
respectively \cite{Stenull1999}. The exponent $\zeta$ describes
the divergence of the resistance of the links when approaching the
gel point. It was previously assumed \cite{DeGennes1976,Sahimi1983}
that $\zeta$ is close to one for $d=3$, which has been confirmed
by more recent renormalization group estimates that yield \cite{Stenull1999}
$\zeta\approx1.05$ and by numerical data \cite{Gingold1990} with
$\zeta\approx1.117\pm0.019$. Literature values of experimental or
simulation data for $\mu$ are largely scattered in the range of \cite{Sahimi1983,Isichenko}
$1.4\le\mu\le2.45$. Our results for $\mu$ range from 1.69 to 1.95
with an average of $1.82\pm0.03$, which is in rather good agreement
with critical percolation, in particular with renormalization group
estimates for all contributing exponents, which yield $\mu\approx1.87$. 

The positions of the gel points are all well above the estimates based
upon intra-molecular reactions, $p_{\text{c,\ensuremath{\mu}}}>p_{\text{c,id}}+\Delta p$,
and do not agree with these within the error of the analysis. Note
also that gel points like $p_{\text{c,\ensuremath{\mu}}}$ are underestimated
systematically in finite samples as shown, for instance, in Figure
1 of \cite{Sahimi1983}. Therefore, our results for $p_{\text{c,\ensuremath{\mu}}}$
are lower bounds for the gel points of macroscopic samples. Since
finite size effects can cause only a systematic enlargement of $\Delta p$
(extra loops through periodic bounds, see the preceding section),
also the observed gaps between $p_{\text{c,\ensuremath{\mu}}}$ and
$p_{\text{c,id}}+\Delta p$ are lower bounds for the equivalent data
of macroscopic samples.

\section*{Gel point estimated based upon weight average molecular weight}

As a second estimate for the gel point, we analyze the weight average
degree of polymerization of the soluble molecules, $N_{\text{w}}$.
Theoretically, a dependence of 
\begin{equation}
N_{\text{w}}\propto\left|\epsilon\right|^{-\gamma}\label{eq:Nw}
\end{equation}
is expected, with best estimates for the exponent of $\gamma\approx1.82\pm0.04$
(3rd order $\epsilon$ expansion \cite{Alcantara1981}), $\gamma=1.805\pm0.02$
(from series expansion of the cluster size distribution \cite{Adler1990}),
or $\gamma=1.795\pm0.005$ (Monte-Carlo data by R. M. Ziff and G.
Stell \cite{Adler1990,Strenski1991,Stenull1999}). Experimental data
fits to exponents in the range of \cite{Isichenko} $1.63\le\gamma\le1.91$.
For mean field, the corresponding exponent is $\gamma=1$. 

\begin{figure}
\includegraphics[angle=270,width=1\columnwidth]{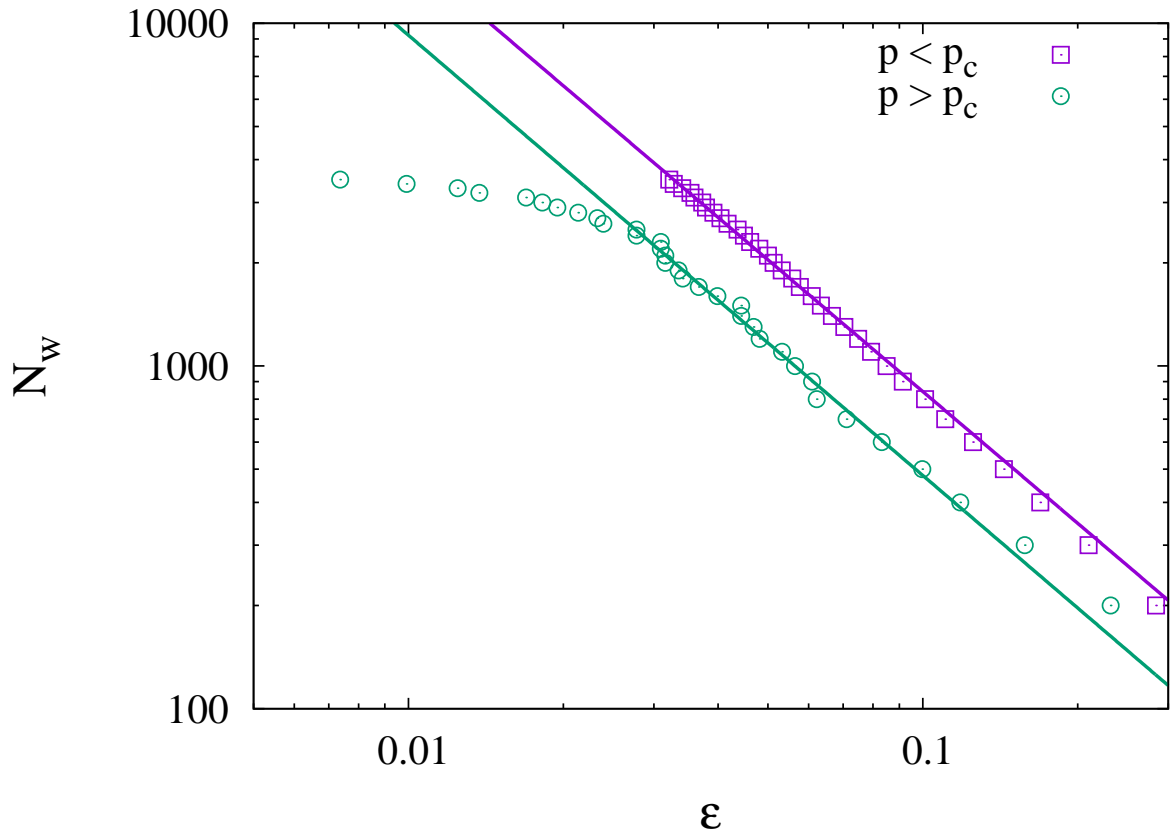}

\caption{\label{fig:Extrapolation}Plot of $N_{\text{w}}$ data as a function
of $\epsilon$ for a determination of $p_{\text{c,\ensuremath{\gamma}}}$
and $\gamma$. The extrapolation of the samples $f=6,$ $N=64$ with
the smallest number of cross-links and largest deviation to the predicted
$\gamma$ is shown.}
\end{figure}

The gel point estimate $p_{\text{c,\ensuremath{\gamma}}}$ and exponent
$\gamma$ is determined from $N_{\text{w}}$ data as suggested in
Refs. \cite{Stauffer1982,Shy1985,Herrmann1986}: we plot simultaneously
the reduced $N_{\text{w}}(p)$ (excluding the largest cluster \cite{Herrmann1986,Lee1990,Cheng1994,Cail2001,Lang2016,Wang2017})
above $p_{\text{c}}$ and the total $N_{\text{w}}(p)$ below $p_{c}$
as a function of $|\epsilon|$ and shift $p_{\text{c}}$ such that
the steepest decay\footnote{For our simulations, there is a cross-over to mean field at large
$|\epsilon|$ (because of large overlap number of junctions) and a
cross-over of the $p>p_{\text{c}}$ branch to saturation due to finite
size at $|\epsilon|\rightarrow0$, which is the reason why only the
steepest decay can become linear on a log-log plot.} of both branches shows the same slope. The resulting estimate for
$p_{\text{c}}$ is denoted by $p_{\text{c,\ensuremath{\gamma}}}$
for a better distinction from other estimates and summarized in Table
\ref{tab:The-critical-conversion}. The exponent $\gamma$ including
the error estimate from the fit is given also in Table \ref{tab:The-critical-conversion}.
$p_{\text{c,\ensuremath{\gamma}}}$ had to be adjusted with four digits
accuracy to achieve parallel upper and lower branches of $N_{\text{w}}(p)$,
however, there is typically less than one order of magnitude in $|\epsilon|$
where the $N_{\text{w}}$ data appears to be linear on a log-log plot,
see Figure \ref{fig:Extrapolation} for the ``worst case'' example
where $\gamma$ deviates most from the prediction of critical percolation.
In comparison, the modulus data is linear on a log-log plot for typically
one and a half decades. Therefore, we expect that $p_{\text{c,\ensuremath{\mu}}}$
should be more accurate than $p_{\text{c,\ensuremath{\gamma}}}$ for
our data in contrast to typical percolation studies, where $p_{\text{c,\ensuremath{\gamma}}}$
is considered as the best estimate \cite{Stauffer1982}. Accordingly,
we expect also a somewhat larger error for $p_{\text{c,\ensuremath{\gamma}}}$
of $\gtrsim10^{-3}$ as compared to $p_{\text{c,\ensuremath{\mu}}}$
in contrast to the four digits accuracy to parallelize the steepest
decays of both branches of $N_{\text{w}}$.

The average difference between both gel point estimates is $p_{\text{c,\ensuremath{\mu}}}-p_{\text{c,\ensuremath{\gamma}}}=0.0012\pm0.004$
and thus, below the cumulated error of both estimates, which confirms
the consistency of our analysis. Note that this difference is about
one order of magnitude smaller than the observed smallest gap between
$p_{\text{c,id}}+\Delta p$ and our estimates $p_{\text{c,\ensuremath{\mu}}}$
or $p_{\text{c,\ensuremath{\gamma}}}$. Therefore, the observation
of this gap is significant and we present below an analysis of the
scaling of the gel point based upon $p_{\text{c,\ensuremath{\mu}}}$
that we consider as our best estimate. Nevertheless, an additional
test using $p_{\text{c,\ensuremath{\gamma}}}$ is included in Figure
\ref{fig:Gel-point-shift} and \ref{fig:Gel-point-shift-1}, which
confirms in each case the observed scaling of $p_{\text{c,\ensuremath{\mu}}}$
within error bars.

For our finite simulations, a $\gamma$ between mean field and critical
percolation is found, with the tendency that smaller samples and samples
with a higher overlap number are closer to the mean field prediction.
This is a typical observation for finite samples within medium-range
bond percolation problem \cite{Ouyang2018}, since the number of independent
volumes in one direction scales as $L/R$, where $L$ is the lattice
size and $R$ the range of the bonds. For $L/R=1$, one reaches the
mean field limit as bonds can be introduced between any pair of nodes,
while for $R\approx bN^{1/3}/\phi$, one arrives at the classical
bond percolation limit. Our simulation data spans only a rather small
range of $L/R$ and the error for $\gamma$ is rather large such that
a detailed analysis of the cross-over scaling is not much meaningful. 

The ratio of the coefficients of the two power law fits below and
above the gel point, $C_{-}/C_{+}$, is also provided in Table \ref{tab:The-critical-conversion}.
For this ratio, a value of $\approx10$ is expected for percolation
in three dimensions while kinetic gelation provides $C_{-}/C_{+}\approx3$
with the possibility that $C_{-}/C_{+}\rightarrow1$ in the limit
of mean field (diverging overlap number) \cite{Herrmann1986}. A similar
tendency is found for the ratio $C_{-}/C_{+}$ that tends towards
unity when $\gamma$ approaches one while it is largest for the samples
with largest $\gamma$. However, a more detailed analysis of $C_{-}/C_{+}$
as for the cross-over scaling of the exponents is not possible within
the limited data of the present study and subject of ongoing research.

\section*{Scaling of the gel point shift}

\begin{figure}
\includegraphics[angle=270,width=1\columnwidth]{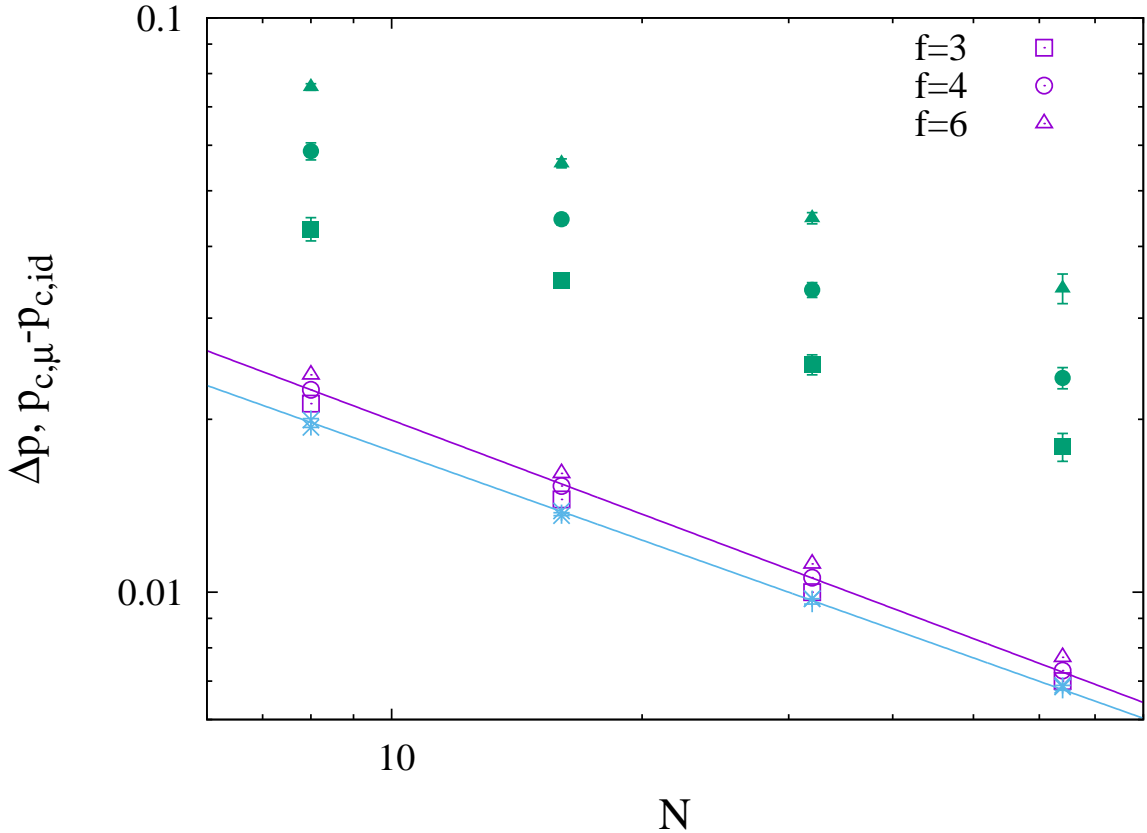}

\caption{\label{fig:Collapse1}Gel point shift as obtained from modulus, $p_{\text{c,\ensuremath{\mu}}}-p_{\text{c,id}}$
(full symbols) compared with $\Delta p$ as measured at the peak of
$N_{\text{w}}$ (open symbols) or in self-consistent manner (plus,
times, and star symbols for $f=3$, 4, and 6 respectively) at a conversion
of $p_{\text{c,id}}+\Delta p$. The upper line is a power law decay
for $\ensuremath{\Delta p=}cN^{-\alpha}$ with $c=0.070\pm0.004$
and $\alpha=0.54\pm0.03$, the lower line a similar power law decay
with $c=0.057\pm0.001$ and $\alpha=0.52\pm0.01$.}
\end{figure}

The first goal of the present paper is to demonstrate that the true
gel point does not overlap with estimates based upon intra-molecular
reactions. Therefore, we take $p_{\text{c,\ensuremath{\mu}}}$ as
our best estimate for the true gel point and compare with the upper
bound for a mean field estimate with a correction for cyclization,
$p_{\text{c,id}}+\Delta p$ in Figure \ref{fig:Collapse1}. The data
are plotted as a function of $N$ in order to check whether the shift
of the gel point scales with the overlap number of elastic strands
in melt as proposed in mean field models. This is observed indeed
for the $\Delta p$ data but not for the true gel point shift $p_{\text{c,\ensuremath{\mu}}}-p_{\text{c,id}}$.
According to the above discussion and section ``intra-molecular reactions''
of the Appendix, it is expected that the amount of intra-molecular
reactions is independent of $f$ in a very good approximation, which
is demonstrated by the excellent collapse of the data for the self-consistent
determination of $\Delta p$ in Figure \ref{fig:Collapse1}. This
is in line with older models for the gel point shift, for instance
Refs. \cite{Kilb1958,Ahmad1980,Suematsu2002} but disagrees with more
subsequent work, for instance Refs. \cite{Rolfes1993,Wang2017}.

\begin{figure}
\includegraphics[angle=270,width=1\columnwidth]{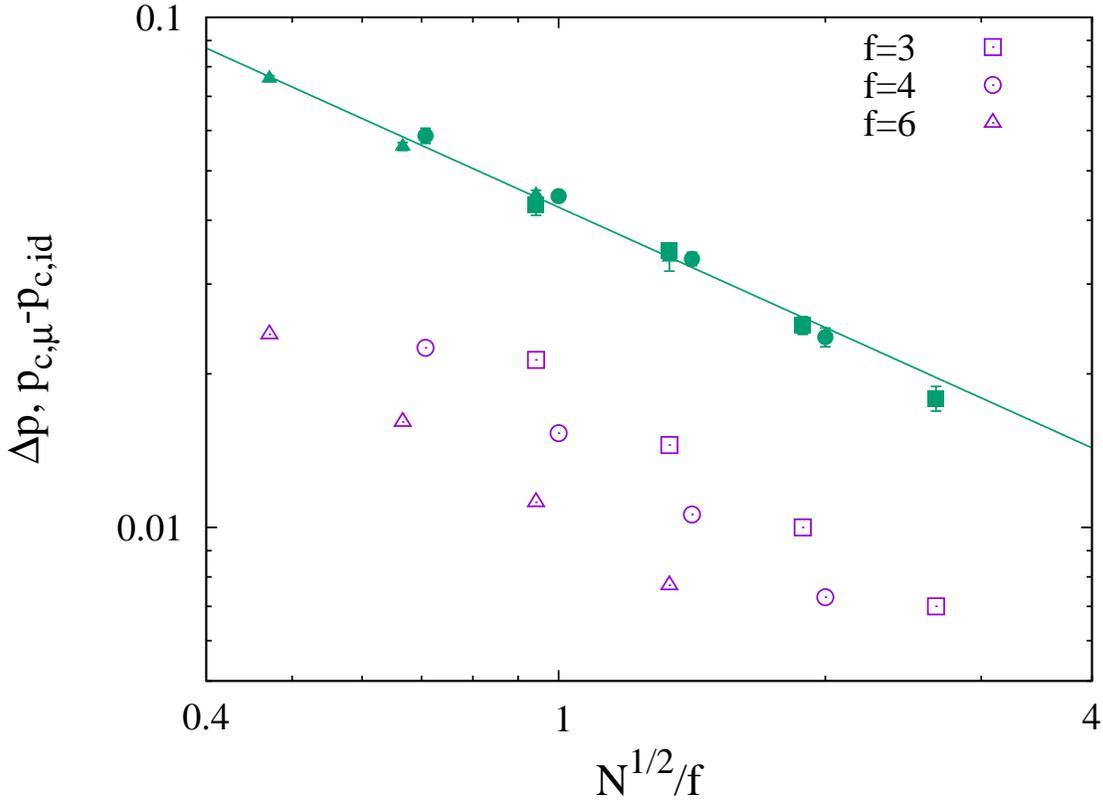}

\caption{\label{fig:Collapse2}Gel point shift as obtained from extrapolation
of modulus data, $p_{\text{c,\ensuremath{\mu}}}-p_{\text{c,id}}$
(full symbols) compared with $\Delta p$ as measured at the peak of
$N_{\text{w}}$ (open symbols). The line is a power law decay $c\left(N^{1/2}/f\right){}^{-\alpha}$
with $c=0.0424\pm0.0005$ and $\alpha=0.78\pm0.03$.}
\end{figure}

The second interesting observation of Figure \ref{fig:Collapse1}
is that $p_{\text{c,\ensuremath{\mu}}}-p_{\text{c,id}}$ is neither
in quantitative nor in qualitative agreement with $\Delta p$ and
is indeed an explicit function of $f$ (similar results are found
for $p_{\text{c,\ensuremath{\gamma}}}-p_{\text{c,id}}$). In a recent
paper on a two-dimensional long range percolation problem, it was
shown that the shift of the gel point with respect to mean field (for
small deviations from mean field) is a power law function of the number
of accessible neighbors \cite{Ouyang2018}. We expect a similar behavior
in three dimensions based upon the overlap number of sites (cross-links)
$P$ . Indeed, a collapse of the $p_{\text{c,\ensuremath{\mu}}}-p_{\text{c,id}}$
data are obtained for a scaling variable $P\propto N^{1/2}/f$ for
our cross-linked melts, see Figure \ref{fig:Collapse2}.

An overlap of the data as a function of the overlap number of chains
$\propto N^{1/2}$ has been emphasized previously \cite{Wang2017}
and was proposed in other models on the gel point shift \cite{Kilb1958,Ahmad1980,Suematsu1998,Cail2007}
due to the expected dependence of $\Delta p$ on $N^{1/2}$ as discussed
above. However in Ref. \cite{Wang2017}, only data with $f=4$ were
analyzed. Therefore we compare in section ``Experimental gel point
data'' of the Appendix the gel point data of several studies for
networks made of similar chemistry but different $f$. Unfortunately,
the available experimental data are not fully conclusive, since not
all data show exactly the same trend. Nevertheless, the common result
is that any set of data exhibits clear power law dependence $P^{-\alpha}$
for the observed delay of the gel point. The data for the PDMS systems
in \cite{Cail2007} is in excellent agreement with the proposed dependence
on $f$, while there is only fair agreement in this respect for the
PU systems as discussed in the Appendix. Cail and Stepto \cite{Cail2007}
point out that the mean field prediction for the amount of loops underestimates
the real shift of the gel point by a factor of four to seven (a factor
of two to five for our simulation data). Therefore, neither for these
experimental data nor for our simulations, there is a quantitative
match of the gel point shift with the amount of intra-molecular reactions
at the gel point.

In general, there is $P\propto\phi N^{1/2}/f\propto\left(V_{\text{dry}}/V\right)$
for networks prepared in semi-dilute solutions with theta solvents
and $P\propto\left(V_{\text{dry}}/V\right)^{0.65}$ in good solvents
using standard scaling relations for $R$ in both cases \cite{Rubinstein2003}.
Here, we used $V_{\text{dry}}$ for the ``dry'' volume of the polymers
and $V$ for the volume of the solution at preparation conditions.
Note that the scaling $\propto N^{1/2}$ is only asymptotically correct,
since corrections to polymer size due to an incompletely screened
excluded volume \cite{Lang2015a} are neglected above. Furthermore,
we have to restrict our discussion to the case of $P>1$ in order
to avoid a significant impact of diffusion as discussed in the introduction.
A fit of the data shows that the shift of the gel point scales as
$P^{-\alpha}$ with a power $\alpha\approx0.78\pm0.03$. Note that
data at the same $P/f$ but different $f$ refers to samples containing
a different number of junctions, since all samples were simulated
in a box of size $L=256$ lattice units. Since these data superimpose
for our scaling variable, we do not expect a dramatic effect of finite
size corrections for our gel point estimates. 

\begin{figure}
\includegraphics[angle=270,width=1\columnwidth]{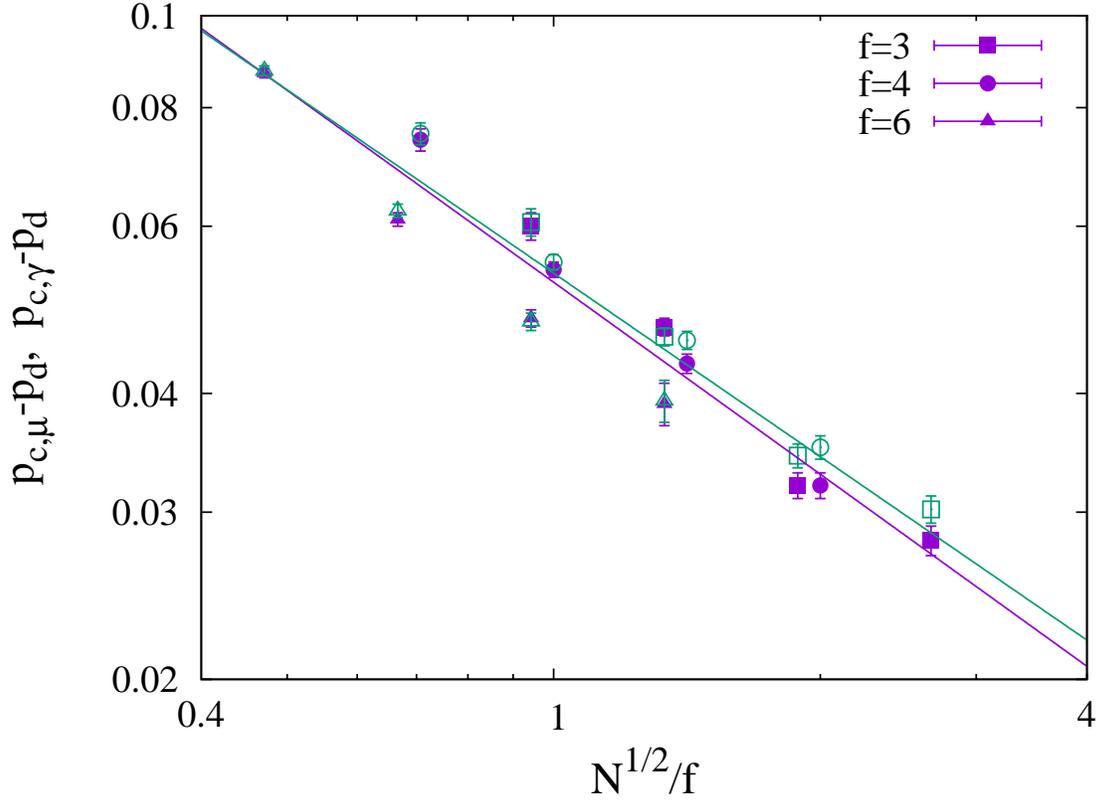}

\caption{\label{fig:Gel-point-shift}Gel point shift as obtained from modulus
data, $p_{\text{c,\ensuremath{\mu}}}-p_{\text{d}}$ (full symbols)
compared with gel point shift from peak of $N_{\text{w}}$ data, $p_{\text{c,\ensuremath{\gamma}}}-p_{\text{d}}$
(open symbols). The lines are power law decays $c\left(N^{1/2}/f\right){}^{-\alpha}$
with $c=0.052\pm0.001$ and $\alpha=0.67\pm0.06$ (for modulus data)
and $c=0.054\pm0.001$ and $\alpha=0.64\pm0.06$ for the $N_{\text{w}}$
data.}
\end{figure}

The above analysis is equivalent to raw experimental data where no
corrections to scaling have been considered. Figure \ref{fig:Gel-point-shift}
contains a more accurate analysis that corrects for unequal reactivity
and composition fluctuations by relating the true gel point to $p_{\text{d}}$
instead of $p_{\text{c,id}}$. On experimental side, a similar analysis
could be conducted using network disassembly spectrometry (NDS) \cite{Zhou2012}
next to the critical point. As a result, we obtain a slightly smaller
$\alpha$ close to $2/3$. Recall that $p_{\text{d}}<p_{\text{c,id}}$
for our samples, since the corrections are dominated by composition
fluctuations. Therefore, a larger $\alpha$ from experimental data
on end-linking of model networks as compared to an apparent $\alpha\approx2/3$
can be taken as a possible indication of imperfect mixing. Indeed,
this discussion is supported by experimental studies, see section
``Experimental gel point data'' of the Appendix for more details.
In brief, the exponents $\alpha$ that one obtains from the data of
Refs. \cite{Gordon1967,Spouge1986,Cail2007,Tanaka2012,Nishi2017}
as a function of the chain overlap number ranges from $0.63$ up to
$1.36$. In several cases where large exponents $\alpha>1$ are fit
to the experimental data, the authors of the original works mention
systematic deviations from other studies and suspect unequal reactivity
or mixing problems to be the reason for these changes.

So far, we have only considered the total delay of the gel point.
Clearly, if one introduces a new bond into the system, this bond either
connects two nodes of the same cluster or two nodes of different clusters.
Thus, the contributions of intra-molecular reactions and XB to the
gel point delay are disjoint and we can single out the XB contribution
by analyzing $p_{\text{c}}-\left(p_{\text{d}}+\Delta p\right)$. In
this respect, we have to mention that intra-molecular reactions are
by far more important for GS as compared to percolation when comparing
data for systems with the same number of neighbors. The reason for
this difference is that loops with $i\le2$ can be formed in GS, which
are typically forbidden in bond percolation models. Since the size
distribution of loops decays quickly as $i^{-5/2}$, this yields almost
one order of magnitude more intra-molecular reactions for GS as compared
to percolation problems with a similar number of accessible neighbors.
Thus, if XB is the main source for a shift of the critical point away
from the mean field gel point in percolation problems, one must not
neccessarily expect the same for polymer gelation.

\begin{figure}
\includegraphics[angle=270,width=1\columnwidth]{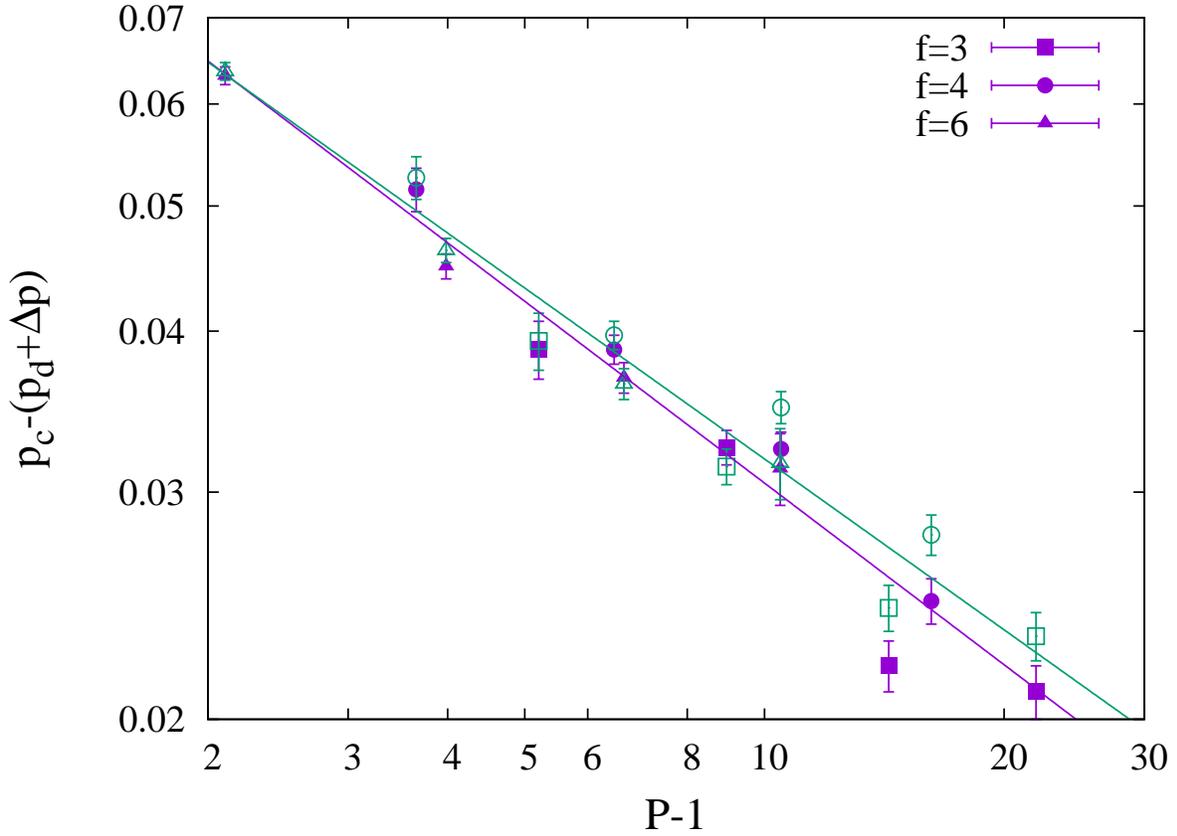}

\caption{\label{fig:Gel-point-shift-1}XB contribution to the gel point shift
as obtained from either modulus data (full symbols) or $N_{\text{w}}$
data (open symbols) with corrections for chain size and unequal reactivity.
The lines are power law decays $c\left(P-1\right){}^{-\alpha}$ with
$c=0.090\pm0.004$ and $\alpha=0.47\pm0.03$ (for modulus data) and
$c=0.088\pm0.004$ and $\alpha=0.44\pm0.03$ for the $N_{\text{w}}$
data.}
\end{figure}

Above, we have not accounted for long range bond correlations that
lead to an apparent swelling of chain size \cite{Semenov1996,Wittmer2007,Lang2015a}.
This can be corrected by including more accurate estimates for the
size of the chain end-to-end vector $R$ in equation (\ref{eq:Pj}),
see for instance, equation (7) of Ref. \cite{Lang2015a} with the
corresponding corrections for our simulation model. In effect, the
consideration of non-ideal chain conformations stretches and shifts
the data along the $x$-axis, which causes a smaller exponent $\alpha$.
A qualitatively similar correction is obtained when subtracting $\Delta p$,
since $\Delta p$ decays quicker than $p_{\text{c}}-p_{\text{d}}$,
see Figure \ref{fig:Collapse1}. If both corrections are applied,
the exponent is reduced from $\alpha\approx2/3$ to $\alpha\approx0.56$.
Further reduction of the exponent results from plotting the data as
a function of the number of accessible neighbors, $P-1$, in order
to introduce the same abscissa as in percolation studies ($P-1$ is
the equivalent of the coordination number of the lattice). These modifications
lead in total to $\alpha\approx0.47$ (based upon modulus data), see
Figure \ref{fig:Gel-point-shift-1}. Note that similar corrections
can be included when analyzing experimental data, since the size of
the molecules can be measured and $\Delta p$ is also accessible through
NDS \cite{Zhou2012}.

Figure \ref{fig:Gel-point-shift-1} still does not indicate a large
systematic shift of the $p_{\text{c}}-\left(p_{\text{d}}+\Delta p\right)$
data for samples with a different number of nodes (data at same $P$
but for different $f$). This supports our proposal that finite size
corrections will not lead to a dramatic modification of $\alpha$.
A more detailed finite size analysis with an exact determination of
the exponent $\alpha$ requires samples of a range of different sizes.
For this purpose, we currently develop a specialized simulation approach
for our particular bond percolation problem as part of ongoing work.
Quantitatively, we expect a weak increase of $\alpha$, since finite
size effects lead to an underestimation of $p_{\text{c}}$, whereby
this underestimation is larger for samples with a smaller ratio $L/R$
(i. e. larger $N$ and larger $P$). Thus, $\alpha\approx0.47$ should
be considered as a lower bound for the true exponent $\alpha$ of
macroscopic samples.

Below, we devlop a rough idea where one could expect the exponent
$\alpha$ from theoretical side by comparison with recent work on
``long range'' percolation, Ref. \cite{Ouyang2018}. In this work,
$z\sim R^{d}$ is the number of accessible neighbors in the percolation
problem, where $R$ is the range up to which bonds can be formed and
$d$ is the space dimension. Our quantity of interest, $p_{\text{c}}-\left(p_{\text{d}}+\Delta p\right)$,
is approximately $zp_{\text{c}}-1$ in Ref. \cite{Ouyang2018} and
we follow the discussion around Figure 13 in Ref. \cite{Ouyang2018}.
Note that conversion $p$ (and thus, $p_{\text{c}}$) is normalized
in Ref. \cite{Ouyang2018} to the maximum possible number of bonds
per node, which is $z$. In contrast to this, the maximum number of
bonds is $f$ in GS and depends neither on $R$ nor $d$. The second,
more subtle difference between GS and long range percolation is that
the bonds (the polymer molecules) fill space for GS, while the nodes
fill space for the percolation problem. Thus, there is $P\propto1/f$
and both, the range $R$ and the overlap number scale with $N^{1/2}$
in GS. Thus, $P\propto R$ for GS while $z\sim R^{d}$ in percolation
problems.

In order to derive an estimate for $\alpha$, let us assume that for
sufficiently large $f$, the differences between a finite fixed $f$
and a node valence that grows $\propto z$ are not largely relevant
at $p_{\text{c}}$, since $zp_{c}\approx1$ next to the mean field
critical point. Furthermore, $P\propto1/f$ rescales density for GS
by a constant factor and thus, the number of neighbors $P-1$ that
corresponds to $z$ is the key variable for scaling. Then, let us
parametrize the neighborhood of the MF fixed point (the limit of $R\rightarrow\infty$)
with a temperature variable $v_{\text{t}}$ and a range parameter
$v_{\text{r}}$ such that these variables rescale as $v_{\text{t}}(s)=s^{y_{\text{t}}^{*}}v_{\text{t}}$
and $v_{\text{r}}(s)=s^{y_{\text{r}}^{*}}v_{\text{r}}$ under a scale
factor $s$. The exponents $y_{\text{t}}^{*}$ and $y_{\text{r}}^{*}$
are the effective mean field renormalization exponents next to the
mean field fixed point \cite{Ouyang2018}. $y_{\text{t}}^{*}=1$ for
$d=3$ renormalizes the ``temperature scale'' (this exponent is
equivalent to $1/\nu$ where $\nu$ is the exponent that describes
the divergence of the correlation length in percolation problems)
\cite{Ouyang2018} and $y_{\text{r}}^{*}=2/3$ (for $d=3$) \cite{Deng}
renormalizes an inverse interaction range. Note that we have kept
the notation of Ref. \cite{Ouyang2018} for a better comparison with
revious work and readers should not get confused by an index $t$
(temperature plays no role for percolation). In fact, the discussion
in Ref. \cite{Ouyang2018} makes use of general results for the $q$-state
Potts model that were obtained within the Ising class $(q=2)$ where
temperature is the key variable, but apply also for the percolation
problem $(q=1)$, see Ref. \cite{Ouyang2018} for more details. For
classical long range percolation problems, there is $R^{-1}\sim z^{-1/d}$
in the mean field limit with $z\rightarrow\infty$. Therefore, 
\begin{equation}
v_{\text{t}}(s)\propto v_{\text{r}}(s)^{y_{\text{t}}^{*}/y_{\text{r}}^{*}}\propto R^{-3/2}\propto z^{-1/2}\label{eq:vt}
\end{equation}
in 3 dimensions. By analogy, we approximate $p_{\text{c}}-\left(p_{\text{d}}+\Delta p\right)\approx v_{\text{t}}$
(ingoring small finite loop contributions). We further identify $z=P-1$.
Given that this analogy and identification are correct, we obtain
\begin{equation}
p_{\text{c}}-\left(p_{\text{d}}+\Delta p\right)\propto\left(P-1\right)^{-\alpha}\label{eq:scaling}
\end{equation}
with $\alpha=1/2$ as the exponent for the XB contribution (finite
$z$ correction beyond finite loops) to the delay of the gel point
in macroscopic samples.

Our lower bound estimate $\alpha\approx0.47$ agrees well with this
rough estimate even though $f$ and $P$ are far from the asymptotic
limit. Also, the difference between both estimates is sufficiently
small to agree with the proposal of a weak impact of finite size on
$\alpha$ that we made above. However, there is still the possibility
that finite $f$ corrections compensate significant finite size corrections,
which we cannot disentangle based upon our limited set of data. Additional
simulations on the percolation problem that is equivalent to GS will
help to clarify the above proposal of $\alpha=1/2$ and are part of
ongoing work.

One interesting aspect of our discussion is that the width of the
Ginzburg zone grows quicker towards smaller $P$ as compared to the
gel point delay, $\epsilon_{\text{G}}/\left[p_{\text{c}}-\left(p_{\text{d}}+\Delta p\right)\right]\approx P^{-1/6}$.
Such a qualitative trend allows to accommodate the requirement of
extra bonds due to both XB and loop formation within the Ginzburg
zone for a significant range of $P>1$ (several orders of magnitude
in $P$), since the coefficient for $\epsilon_{\text{G}}$ is of order
unity \cite{DeGennes1979}, while the coefficient of XB is $c\approx0.09$.
Therefore, we expect that $\alpha\approx1/2$ may apply for virtually
all experimental data. Loop formation may dominate over XB corrections
only in the limit of $P<1$, see Figure \ref{fig:Gel-point-shift-1},
where the percolation model breaks down anyways. Conversely, we infer
that the width of the Ginzburg zone provides a natural upper bound
for $\alpha$, thus, $\alpha\le2/3$. Since the gel point must be
located within the Ginzburg zone, a cross-over of $\alpha$ to this
upper bound may be enforced at very large $P$. Altogether, our estimate
for $\alpha$, the simulation results at the available limited $P$,
and the known scaling of the Ginzburg zone are not in conflict for
the parameter range of interest, $P>1$.

\section*{Summary}

We have demonstrated that the position of the gel point cannot be
estimated by considering intra-molecular reactions as the only correction
to mean field, even though cyclization is well approximated by mean
field. Instead, the largest contribution to the gel point delay is
due to the extra bonds (XB) required to build the connections between
the non-overlapping giant molecules. Our data and discussion indicate
that this XB contribution decays roughly as $\left(P-1\right)^{-1/2}$
in contrast to the contribution of finite loops, which decays approximately
as $N^{-1/2}\propto f/P$ for large junction overlap number $P$.
Beyond these two main contributions, there are corrections due to
composition fluctuations and unequal reactivity, non-ideal chain size,
and due to the transition from junction overlap number $P$ to number
of neighbors $P-1$. Finite size corrections are likely to contribute
only little to scaling, since we observe no large shift of the different
data at same $P$ but different $f$. Some of the corrections (unequal
reactivity and composition fluctuations) could be avoided by analyzing
the homo-polymerization of $f$-arm star molecules as model systems.
Other corrections like non-ideal chain conformations and contribution
of small cycles are accessible through additional measurements, which
allows to repeat our analysis with experimental data.

\section*{Acknowledgement}

We thank the ZIH Dresden for a generous grant of computation time
and the DFG for funding Project LA2735/5-1. We further thank Y. Deng
for stimulating discussions on medium range percolation.

\section*{Appendix}

\subsection*{Intra-molecular reactions}

Loops (cyclic structures in the reaction bath) are analyzed using
a spanning tree approach \cite{Lang2001b} to assure that no loop
is counted twice and that their total number equals the cycle rank
$\xi$. It has been shown by computer simulations \cite{Lang2005a}
that the frequency $L_{i}$ of finite loops made of $i$ precursor
polymers scales at the gel point approximately as predicted by mean
field \cite{Dusek1978,Suematsu2002}

\begin{equation}
L_{\text{i}}(p_{\text{c}})\propto i^{-5/2}.\label{eq:Li}
\end{equation}
Note that within the mean field approximation, the expected average
number of junctions in generation $i$ apart from a given junction
is exactly unity at the gel point independent of junction functionality
$f$. Thus, also the total amount of loops next to the gel point is
independent of $f$ in a good approximation, see Figure \ref{fig:Collapse1}.
Note that integration over the ideal reaction rate, equation (20)
of Ref. \cite{Lang2005a}, up to the gel point, turns the $i^{-d/2}$
dependence for the return probability of a random walk in $d$ dimensions
to $i^{-d/2-1}$ for the (mean field) loop size distribution at the
gel point. In case of excessive loop formation, the above power law
may not be reached for the smallest $i$, see also equation (A2-44)
of Ref. \cite{Dusek1978}.

\begin{figure}
\includegraphics[angle=270,width=1\columnwidth]{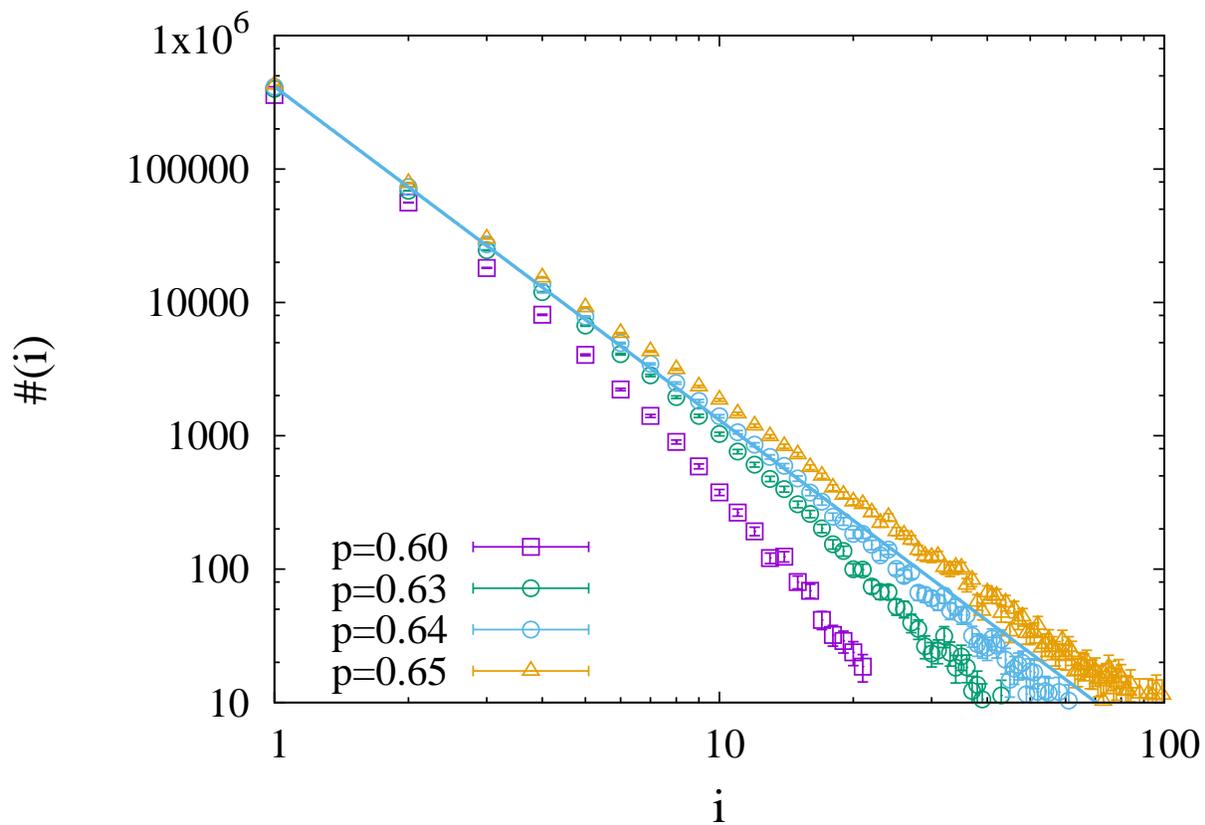}

\caption{\label{fig:size-dist}Total number of loops, $\#(i)$ made of $i$
strands in the vicinity of the gel point for the 100 networks with
$f=4$ and $N=8$. The line indicates a power law decay with exponent
$-5/2$. }
\end{figure}

Loop formation is not excessive for our simulations and the proposed
power law is reached in a good approximation between a conversion
of $p=0.64$ and $p=0.65$, see Figure \ref{fig:size-dist}. But according
to Table \ref{tab:The-critical-conversion}, the data at $p=0.60$
of Figure \ref{fig:size-dist} refers to the gel point as estimated
from intra-molecular reactions, $p_{\text{c,id}}+\Delta p$. In contrast
to this, the loop size distribution at $p=0.60$ is typical for $p<p_{\text{c}}$,
where a cut-off for the size distribution near a finite $i$ reflects
the average number of strands between pairs of reactive groups of
the characteristic molecule, see section ``Mean field estimates of
the gel point''. At the gel point, the degree of polymerization of
the characteristic molecule diverges and thus, the position of the
cut-off shifts towards infinity. In fact, a cross-over to a second
weaker power law at large $i\approx20$ is expected near $p_{\text{c}}$
which is essentially lost in the noise of the data at $i\ge20$.

\subsection*{Unequal reactivity}

Let us consider systems with a stoichiometric ratio $r=1$ of the
reactive groups on chain ends and on junctions. In the mean field
model, the reactivity of all reactive groups is identical. The probability
that an $f$-functional junction has $j$ bonds to chain ends is then
described by the binomial distribution 
\begin{equation}
P(X_{\text{j}})=\left(\begin{array}{c}
f\\
j
\end{array}\right)p^{j}\left(1-p\right)^{f-j}.\label{eq:binom}
\end{equation}
In similar manner, one can compute the connectivity distribution of
chains by considering them as two functional units, $f=2$.

A fully equivalent treatment to this statistical discussion is to
consider reaction rates $k_{\text{j}}\propto f-j$ for the rate equations
\begin{equation}
\frac{\text{d}c_{0}}{\text{d}t}=-k_{0}c_{0}c_{\text{e}}\label{eq:dc1}
\end{equation}
for $j=0$ and 
\begin{equation}
\frac{\text{d}c_{\text{j}}}{\text{d}t}=\left[k_{\text{j}-1}c_{\text{j}-1}-k_{\text{j}}c_{\text{j}}\right]c_{\text{e}}\label{eq:dc2}
\end{equation}
for $0<j<f$ and 
\begin{equation}
\frac{\text{d}c_{\text{f}}}{\text{d}t}=k_{\text{f}-1}c_{\text{f}-1}c_{\text{e}}\label{eq:dc3}
\end{equation}
for $j=f$ (and a similar scheme for chains that are two-functional).
Here, $c_{\text{j}}$ is the concentration of junctions with $j$
bonds, $c_{\text{e}}$ is the concentration of the reaction partners
(the chain ends). By selecting reaction rates $k_{\text{j}}\propto f-j$
one takes into account that there are $f-j$ non-reacted groups per
molecule with same reactivity. For stoichiometric systems, one arrives
after some algebra (and considering $\text{d}p/\text{d}t$) at equation
\ref{eq:binom}.

In case of end-linked model networks, the network junctions are less
accessible to reactions and less mobile with increasing number of
chains attached. Thus, we expect that the reaction rates $k_{\text{j}}$
are no more $\propto f-j$. Instead, there will be comparatively smaller
reaction rates the larger the $j$. Such a behavior leads qualitatively
to a quicker decay of $c_{0}$ and a delay of the increase of $c_{\text{f}}$
as compared to the statistical case, when plotting the corresponding
$P(X_{\text{j}})$ as a function of $p$. Intermediate states $0<j<f$
will exhibit a narrower peak as a function of $p$.

Copolymerizations are inevitably subject to composition fluctuations
of the two species. Composition fluctuations will increase $P(X_{\text{j}})$
where $P(X_{\text{j}})$ is convex and will decrease $P(X_{\text{j}})$
within concave domains as composition fluctuations can be modeled
by a spontaneous split of the system into two domains of somewhat
larger and smaller conversions (similar to the discussion of the phase
behavior of polymer solutions in text books where concentration fluctuations
are considered). On a qualitative basis, we expect for dominating
corrections due to composition fluctuations that $P(X_{\text{0}})$
and $P(X_{\text{f}})$ will be enlarged as compared to the statistical
prediction, since these are convex for all $p$, which is the opposite
trend as expected from accessibility and mobility.

\begin{figure}
\includegraphics[angle=270,width=1\columnwidth]{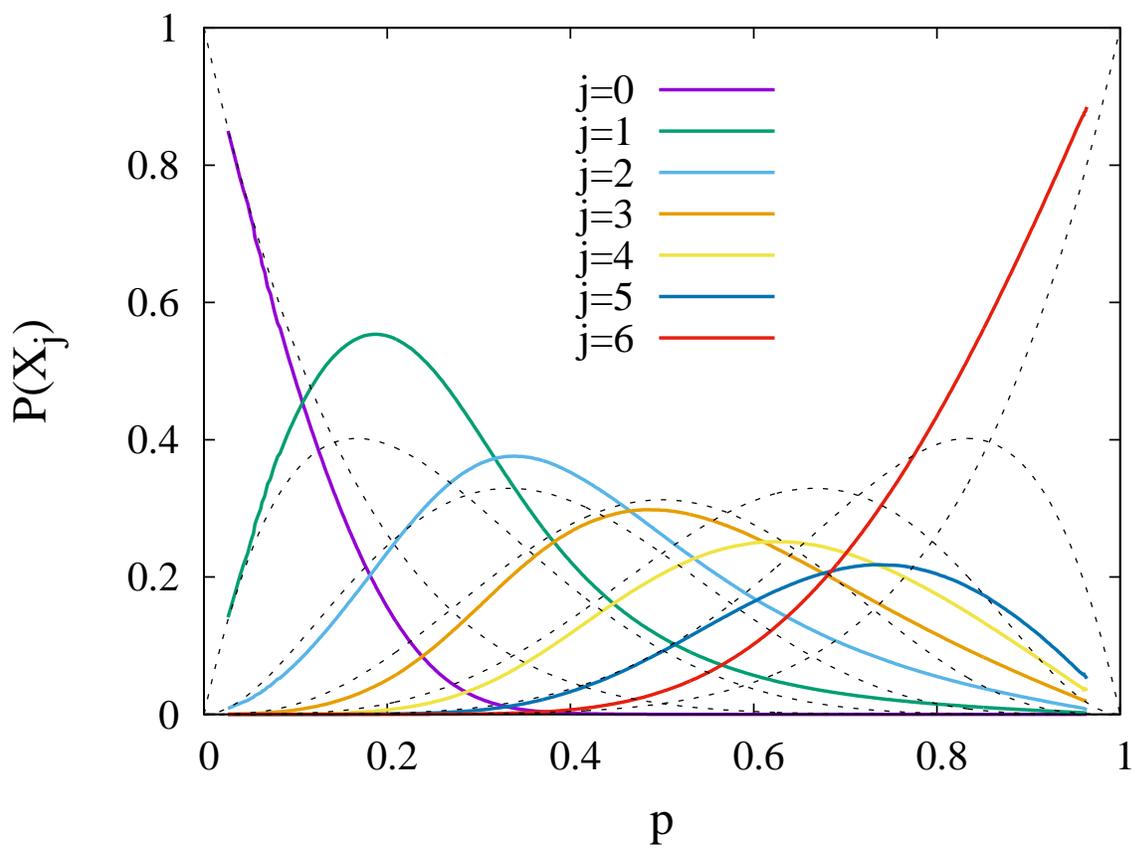}

\caption{\label{fig:Distribution-of-the}Distribution of the number $j$ of
connections to chain ends for junctions $X$ with $f=6$ in networks
with $N=32$. Colored lines show simulation data, dashed black lines
are the binomial distribution, equation (\ref{eq:binom}).}
\end{figure}

\begin{figure}
\includegraphics[angle=270,width=1\columnwidth]{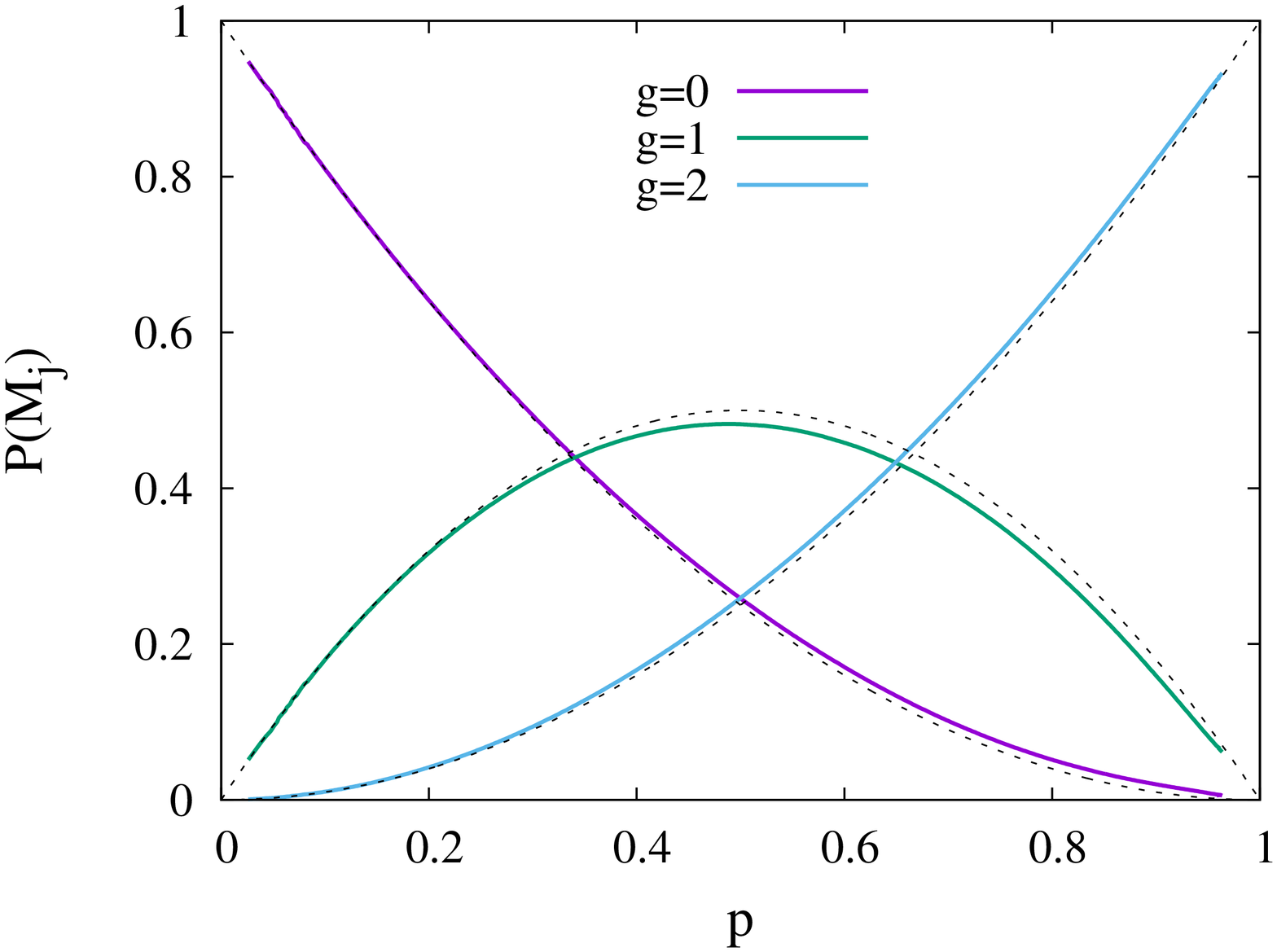}

\caption{\label{fig:Distribution-of-the-1}Distribution of the number $j$
of connections to junctions for chains $M$ in networks with $f=6$
and $N=32$. Colored lines show simulation data, dashed black lines
are the binomial distribution, equation (\ref{eq:binom}).}
\end{figure}

Both effects discussed above are visible in the connectivity distributions
of junctions and chains, see Figure \ref{fig:Distribution-of-the}
and Figure \ref{fig:Distribution-of-the-1}. Junctions are initially
single monomers, which react clearly quicker than junctions attached
to chain ends. Here, the impact of a different mobility dominates.
On the other hand, $P(X_{\text{f}})$ and $P(M_{2})$ appear to be
dominated by the impact of composition fluctuations. Altogether, there
are siginificant deviations from the ideal case, equation (\ref{eq:binom}).
These must be taken into account for an exact mean field estimate
of the critical point.

Non-ideal systems are readily mapped onto the corresponding ideal
systems by considering appropriate distributions of junction and chain
functionality. The experimental data for $P(X_{\text{j}})$ and $P(M_{\text{j}})$
at a given conversion $p$ are considered to be equivalent to a model
system at full conversion with a functionality distribution of junctions
and chains that is identical to $P(X_{\text{j}})$ and $P(M_{\text{j}})$.
Then, equation (24) to equation (33) of Ref. \cite{Macosko1976} can
be used to compute numerically the ``effective'' functionalities
$f_{\text{e}}$ and $g_{\text{e}}$ of junctions and chains, respectively.
The gel point of the equivalent system is given by the condition 
\begin{equation}
(f_{\text{e}}-1)(g_{\text{e}}-1)=1,\label{eq:critical}
\end{equation}
since $rp^{2}=1$ after the mapping of the data and $r=1$ for the
stoichiometric systems of our study. The results for the resulting
estimate of the gel point, $p_{\text{d}}$, based upon this analysis
can be found in Table \ref{tab:The-critical-conversion}. 

\subsection*{Experimental gel point data\label{sec:Experimental-Data-on}}

A large number of authors published experimental data on the gel point
conversion in the past decades starting with the seminal paper by
Flory on gelation \cite{Flory1941a}. Refs. \cite{Gordon1967,Stanford1977,Matejka1980,Spouge1986,Cail2007,Durackova2007,Zhao2006,Duskova2010,Tanaka2012,Nishi2017}
are just some examples from literature. Among these works, we have
chosen Refs. \cite{Gordon1967,Spouge1986,Cail2007,Tanaka2012,Nishi2017}
as basis for comparison, since the data is presented as (or readily
converted into) a function of the so called ``ring forming parameter''
\begin{equation}
\lambda_{\text{0}}=\frac{P_{\text{a}}}{c_{\text{0}}}\label{eq:lambda0}
\end{equation}
that we introduce here only in its simplest possible form for a homo-polymerization
of $f$-functional stars. In this case, $c_{\text{0}}$ is the initial
concentration of reactive groups and $P_{\text{a}}$ is the concentration
of a single reactive group of the star in the vicinity of a selected
reactive group of the same star. Thus, $1/\lambda_{\text{0}}$ is
essentially the overlap number of reactive groups within the pervaded
volume of the shortest strand that can form a loop. Thus, for reactions
in bulk, there is $N^{1/2}/f\propto(\lambda_{0}f)^{-1}$, which allows
for a simple qualitative comparison of simulation data and experiment. 

\begin{figure}
\includegraphics[angle=270,width=1\columnwidth]{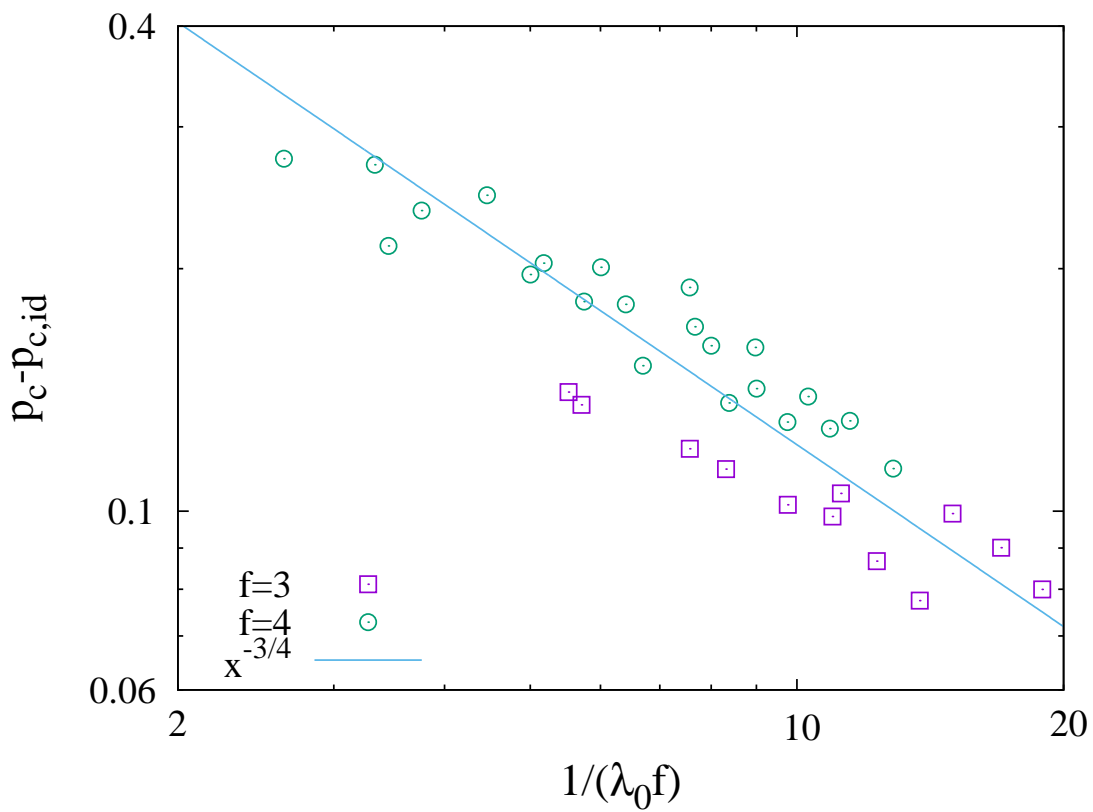}

\caption{\label{fig:Exp1}Collapse of the data for the poly-urethane systems
PU 1-6 of Ref. \cite{Cail2007} that were also discussed in Refs.
\cite{Stanford1977,Stepto1979,Stepto1988,Stepto2000,Cail2001}, where
cross-linking occurred in bulk and in various dilutions in nitrobenzene.
The line indicates a scaling according to $\left(N^{1/2}/f\right)^{-3/4}$.}
\end{figure}

\begin{figure}
\includegraphics[angle=270,width=1\columnwidth]{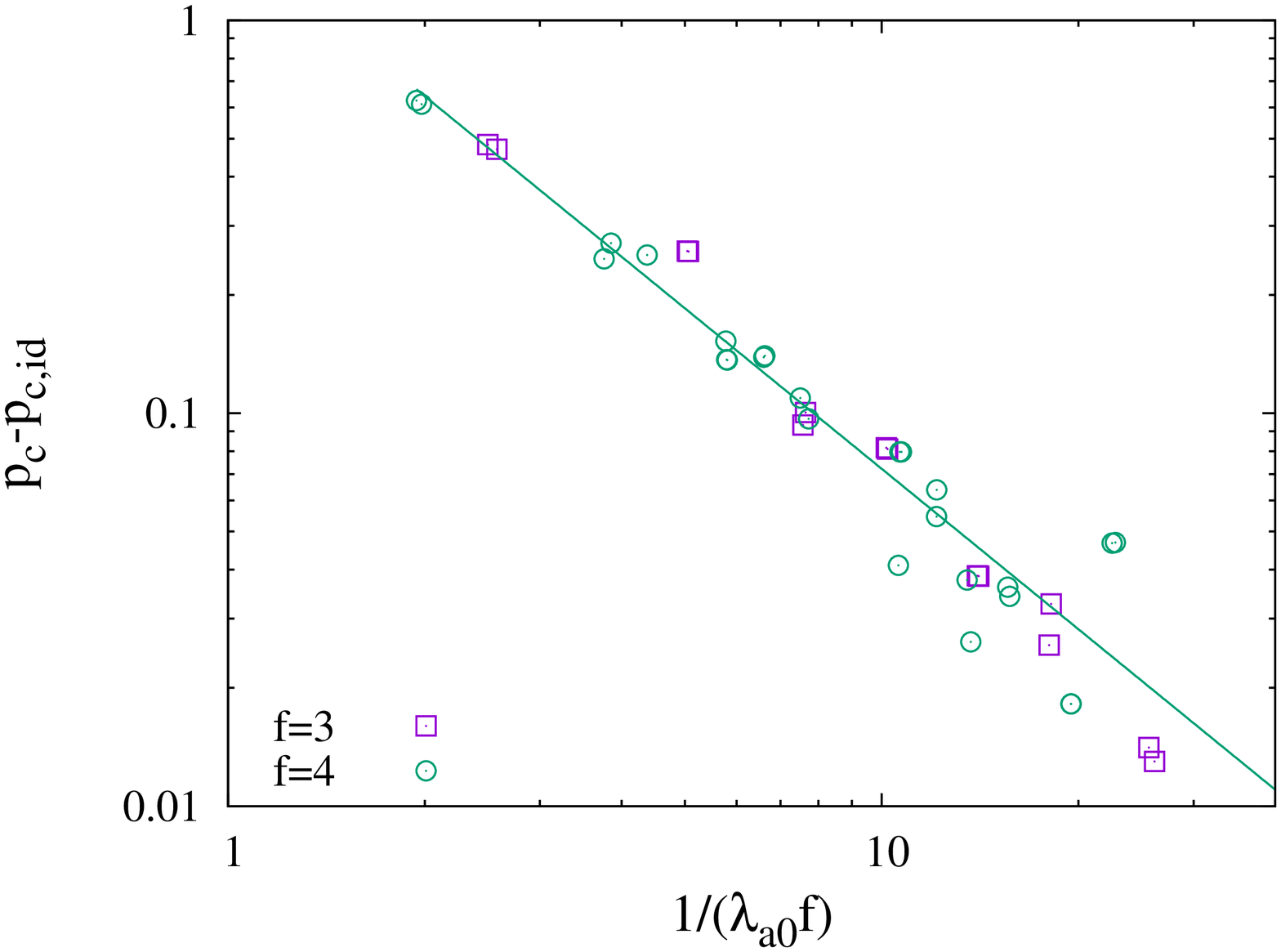}

\caption{\label{fig:Exp2}Collapse of the data for the poly-dimethylsiloxane
systems PDMS 1-4 of Ref. \cite{Cail2007} that were cross-linked in
bulk or diluted by inert PDMS chains. Additional information about
this data can be found in Refs. \cite{Stepto200b,Cail2001}.}
\end{figure}

In Figure \ref{fig:Exp1}, we compare the data of the different PU
systems of Ref. \cite{Cail2007}. The full cloud of data points seems
to support an exponent in the range of $\alpha\approx3/4$ for the
scaling variable $N^{1/2}/f$ while the individual data sets are more
in line with a weaker decay as a function of $N^{1/2}/f$. Furthermore,
the dependence on $f$ is slightly weaker as for our simulation data.
On the other hand, the data of the PDMS systems 1-4, fully overlaps
as a function of $f$, see Figure \ref{fig:Exp2}, but the exponent
for the decay is here $\alpha=1.36\pm0.08$, which is clearly stronger
than for the PU systems or our simulation data. The authors of Ref.
\cite{Cail2007} were aware of these quantitative and qualitative
differences and suspected mixing problems or side reactions to be
responsible for the differences between the experimental data sets.
Such problems were in particular pronounced for PDMS systems 5 and
6, which were not included in the plot for this very reason.

\begin{figure}
\includegraphics[angle=270,width=1\columnwidth]{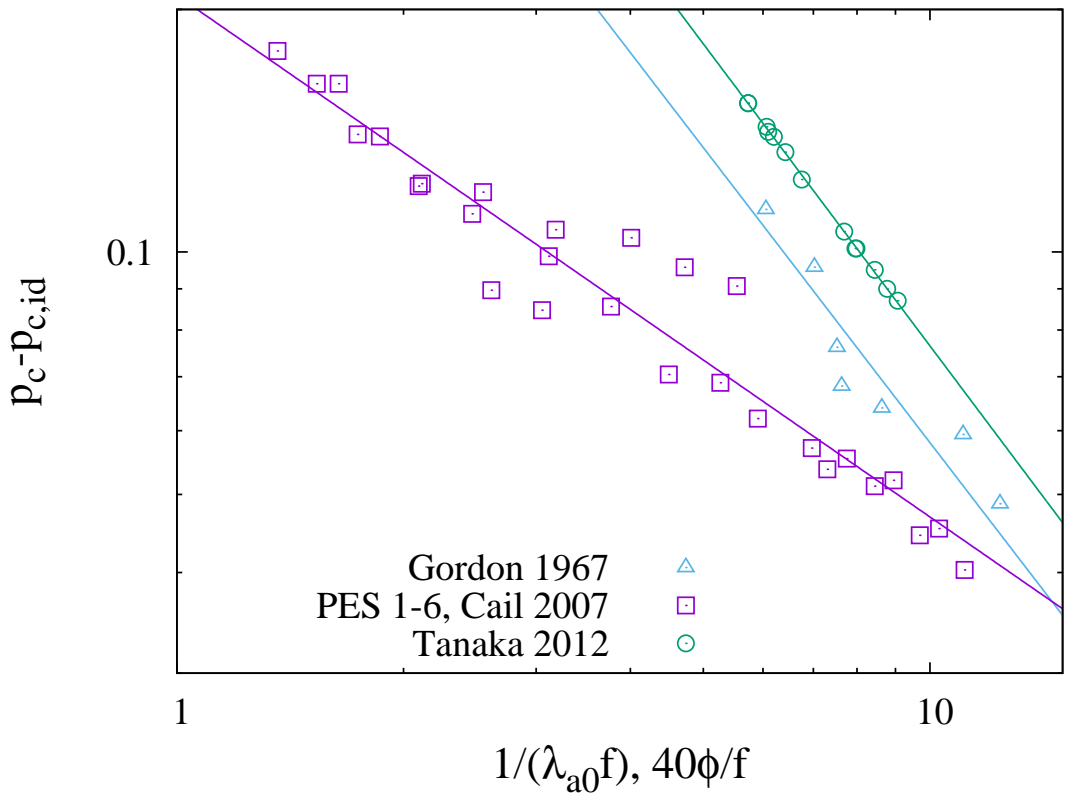}

\caption{\label{fig:Exp3}Gel point data of Cail and Stepto \cite{Cail2007}
on polyester systems PES 1-6 that were reacted in bulk or at various
degrees of dilution with diglyme as solvent (additional information
about this set of data is available in Refs. 1, 2 and 4 of Ref. \cite{Cail2007})
and data by Tanaka\emph{ et al. }\cite{Tanaka2012} where diglycidyl
ether of bisphenol A (DGEBA) reacted with polyoxypropylene (POP) diamine
of different chain lengths at various degrees of dilution with tetraglyme
as solvent. Gordon and Scantlebury \cite{Gordon1967} provide data
on reactions of adipic acid with pentaerythritol either in melt or
with bis(3,6-dioxaheptyl)ether as solvent. Data was analyzed as function
of $\phi$ and shifted by an arbitrary factor of 40 to appear on the
same scale as the data that was analyzed as a function of $\lambda$.}
\end{figure}

Cail and Stepto \cite{Cail2007}, Tanaka et al. \cite{Tanaka2012},
or Gordon and Scantlebury \cite{Gordon1967} report more gel point
data on different series of samples, see Figure \ref{fig:Exp3}. The
observed exponents for the decay are $\alpha=-0.65\pm0.04$, $\alpha=-1.25\pm0.01$,
and $\alpha=-1.2\pm0.2$ respectively, which differ significantly
similar to the PDMS and PU systems discussed above. It is worthwhile
to mention that unequal reactivity seems to be important for the systems
reported by Tanaka et al. \cite{Tanaka2012} where the larger exponent
was recognized, which parallels in part the discussion of the PDMS
systems in Ref. \cite{Cail2007}. Also, Gordon and Scantlebury \cite{Gordon1967}
argue that a substitution effect must occur to explain their set of
data.

\begin{figure}
\includegraphics[angle=270,width=1\columnwidth]{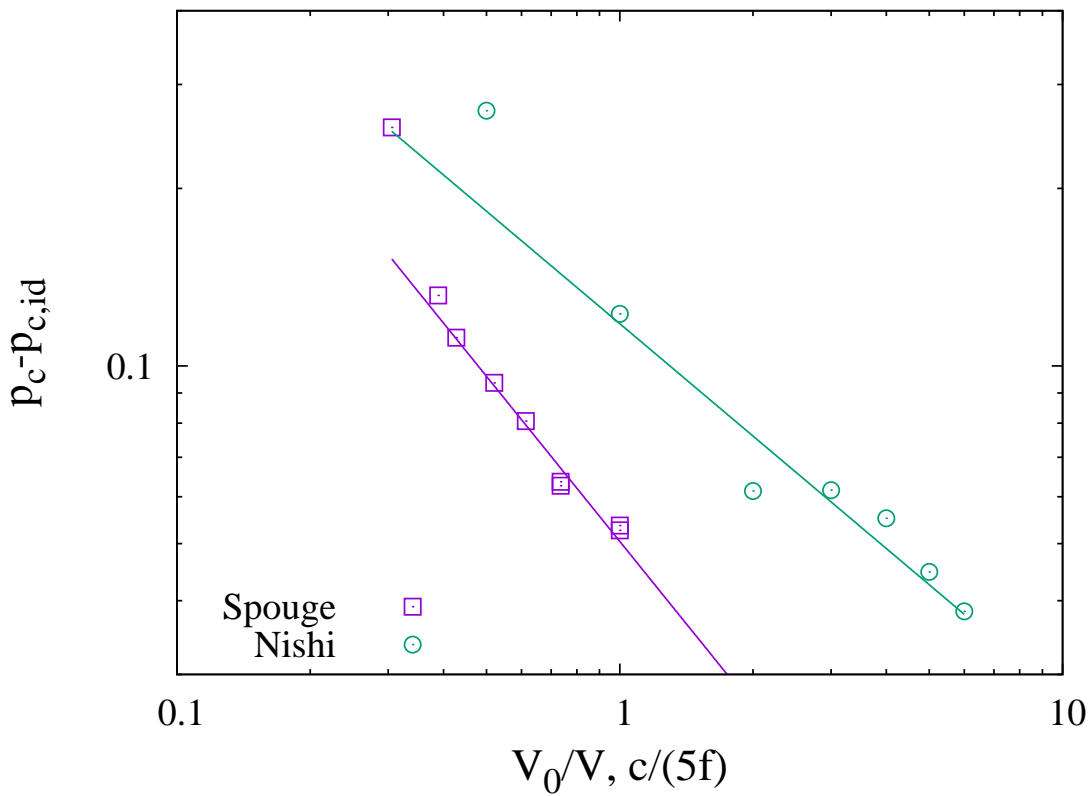}

\caption{\label{fig:Gel-point-data}Gel point data of Spouge \cite{Spouge1986}
on the reaction of adipic acid with pentaerythritol plotted as a function
of $V_{0}/V$ and Nishi \cite{Nishi2017} for the end-linking of 4-functional
stars that is presented as a function of $c/(5f)$ in gram per liter
to fit on the same scale as the data of Spouge.}
\end{figure}

Spouge \cite{Spouge1986} discusses experimental data on the reaction
of adipic acid with pentaerythritol taken from the thesis of Wile
\cite{Wile1945}. When plotting this data as a function of $V_{0}/V$,
an exponent of $-0.93\pm0.05$ is found when excluding the data at
the largest degree of dilution (close to critical dilution), see Figure
\ref{fig:Gel-point-data}. Recently, Nishi et al. \cite{Nishi2017}
report gel point conversions of a copolymerization of 4-functional
stars in the inset of their Figure 3, which we have plotted also in
Figure \ref{fig:Gel-point-data}. This particular type of reaction
suppresses loops made of an odd number of elastic strands \cite{Lang2019},
and thus, there are no pending loops present in these networks. Nevertheless,
there is still a significant delay of the gel point with respect to
the ideal gel point at $1/(f-1)$. The data of Nishi fits to a power
of $-0.63\pm0.08$ when excluding the data point at the lowest concentration
(close to critical dilution) from the fit. An effect of unequal reactivity
can be ignored for the end-linking of stars where all ends of the
arms have the same chemistry and react only once. These exponents
are in the range of our simulation data with $\alpha\approx2/3$ when
using $p_{\text{d}}$ to correct for unequal reactivity and not far
from the PU and the PES systems of Ref. \cite{Cail2007} with observed
exponents for these systems of $\alpha=-0.65\pm0.04$ and $\alpha\approx-3/4$
respectively. Altogether, the PU and the PES systems of Ref. \cite{Cail2007},
the data by Nishi et al \cite{Nishi2017}, and the data of Spouge
\cite{Spouge1986} seem to be less affected by unequal reactivity
and show similar trends as our simulation data.

\subsection*{Numerical Studies in Literature}

The available studies can be split into two major categories depending
on whether a) a mean field model is in the core of the analysis and
the coordinates of all molecules are disregarded or b) the positions
of the molecules in space are taken into account. Our list of examples
from literature below is certainly not exhaustive and focuses on polymer
specific work. Nevertheless, it should allow for a first idea about
the state of the art.

Gordon and Scantlebury \cite{Gordon1967} use a vector of probability
generating functions where loop formation is built in through a set
of differential equations in the coefficients of the vector. The numerical
solutions of their model are used to interpret their experimental
data. This work was criticized later \cite{Sarmoria2001}, since it
does not allow to fit gel point and molecular weight data with the
same adjustable parameters.

The kinetic simulation method developed by Somvarsky and Dusek \cite{Somvarsky1994}
is mean field enriched with Monte-Carlo with an explicit description
of cyclization. This allows to obtain higher order molecular weights
like $N_{\text{w}}$. The data of an early publication were quite
noisy \cite{Somvarsky1994_II} while a later work \cite{Somvarsky1998}
contains gel point data that were obtained under a bit artificial
model for intra-molecular reactions and steric hindrance. These results
are insightful on a qualitative basis, however, a quantitative analysis
remains difficult as there is no obvious choice for all exponents.

Only smallest loops were considered in Ref. \cite{Sarmoria1986}.
Rankin et al. \cite{Rankin2000} compute numerically the gel point
shift from a mean field approach where only three membered rings are
allowed. In both cases, the comparability with real systems that exhibit
a broad distribution of loop sizes is limited.

Pereda et al. \cite{Pereda2001} use rate equations and a large set
of different structures to model the effect of cyclization and unequal
reaction rates on gelation. The numerical results show that the combination
of both effects may either reinforce or partially compensate each
other (depending on the choice of the parameters) even though both
alone lead to a delay of the gel point. 

The most recent work of the mean field class by Wang et al. \cite{Wang2017}
claims a quantitative prediction of gel points based upon Monte-Carlo
simulation where loop formation is introduced randomly into the growing
branched molecules by considering relative contact probabilities inside
reactive groups along the molecular structure with a homogeneous (mean
field) background of contacts with other molecules. It was found that
the frequency of loops decays roughly as a power law $i^{-5/2}$ in
the vicinity of the gel point, where $i$ is the number of linear
strands per loop. The observed power changes from 2.44 to 2.73 with
increasing amount of pending loops and thus, seems to be non-universal,
however, such a qualitative change is expected for small $i$ in case
of excessive loop formation as mentioned in the section ``intra-molecular
reactions'' above. The delay of the gel point was shown to be a function
of the overlap number of elastic strands \cite{Wang2017} and a theoretical
result was obtained where $p_{\text{c}}$ is given as an explicit
function of junction functionality $f$. However, the dependence on
$f$ was not tested since the model was compared only with experimental
data for $f=4$.

Polymer specific simulations on gelation that consider the coordinates
of the molecules in space were pioneered by Leung and Eichinger \cite{Leung}.
The paper by Shy and Eichinger \cite{Shy1985} uses this approach
to estimate the position of the gel points. However, the data is not
consistent as some gel points for large molecular weight lie below
the Flory-Stockmayer prediction despite of a significant amount of
intra-molecular reactions and the simulations do not enforce unequal
reactivity. Also, the intra-molecular reactions show some unexpected
trends (do not extrapolate towards zero for zero conversion; there
is more cyclization for $f=4$ as compared to $f=3$ at low molecular
weight but less at high molecular weight).

\begin{figure}
\includegraphics[angle=270,width=1\columnwidth]{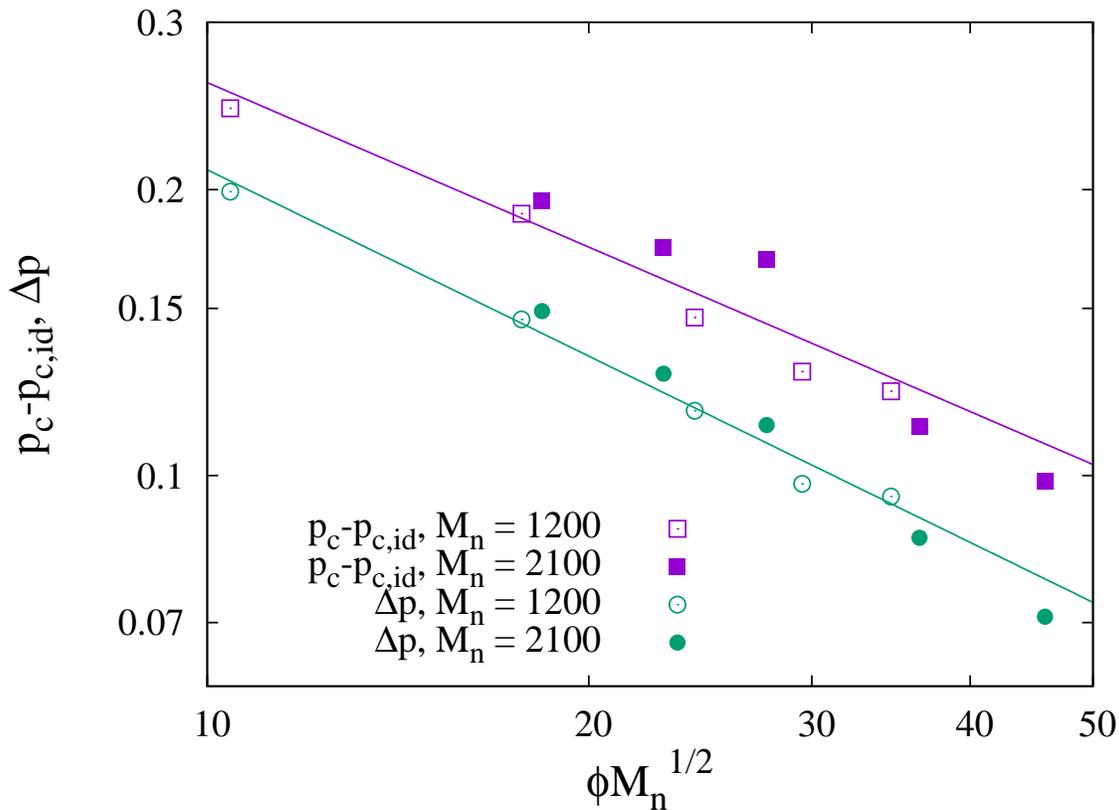}

\caption{\label{fig:Simulation-data-of}Simulation data of Ref. \cite{Lee1990}
for the shift of gel point $p_{\text{c,sim}}-p_{\text{c,id}}$ and
fraction of loops at the gel point, $\Delta p$. }
\end{figure}

In a subsequent work by Lee and Eichinger \cite{Lee1990}, these apparant
problems were resolved, however, the definition of conversion seems
to be based upon the number of connections between two polyoxypropylene
tetrols through one hexamethyleen diisocyanate as obvious from equation
(8), the results for $\lambda_{\text{AB}}$, or the last two columns
of table 3 in their paper (in effect $p=q^{2}$ is used as ``conversion''
where $q$ is the true conversion of reactive groups). With this conjecture,
all of the presented data becomes self-consistent, agrees to the presented
experimental data of Stepto's group, and shows similar qualitative
trends as the data of our work. It is discussed in Ref. \cite{Lee1990}
(independent of this conjecture) that the intra-molecular reactions
are not sufficient to account for the shift of the gel point. Both,
the shift of the gel point and the fraction of intra-molecular reactions
are in accord with a power law decay as a function of the overlap
of the molecules, see Figure \ref{fig:Simulation-data-of}. The corresponding
exponents are $-0.57\pm0.06$ and $-0.65\pm0.03$ respectively, which
are both smaller than $\approx2/3$ and $-1.08\pm0.06$, which were
the corresponding exponents of our data, if plotted with the same
axes and the same level of corrections as Figure \ref{fig:Simulation-data-of}.

Simulation data using the model of Ref. \cite{Leung} was compared
with experimental data of the Stepto group in Ref. \cite{Dutton1994}.
This comparison yields reasonable predictions of gel points from simulation
data, while mean field approaches underestimate systematically the
gel point shift of experimental data.

Gupta et al. \cite{Gupta1991} introduce a long range percolation
model but discuss it as a model for diffusion controlled reactions.
The authors observe that cylization accounts only for a part of the
delay of the gel point. A subsequent work by Hendrickson et al. \cite{Hendrickson1995}
uses an improved version of the original algorithm. Figure 6 of Ref.
\cite{Hendrickson1995} shows that the shift in the conversion is
significant and dominated by intra-molecular reactions for short range
interactions, while for reactions with a longer range (larger overlap
number), the spatial arrangement of the molecules becomes the dominant
cause for the shift of the gel point. This latter contribution grows
sublinear with the overlap number, which agrees qualitatively with
our results. However, one has to be cautious regarding their results
since the criterion used to determine the gel point does not reflect
the true gel point of these systems. Another caveat regards the point
that $f$ decouples from node density in their simulations, while
it does not for real reactions in bulk, which leads to some counterintuitive
results for the dependence on $f$.

Yang et al. \cite{Yang2007} study network formation using Molecular
Dynamics simulations. Apparently, not all of the presented data seem
to be consistent and thus, need to be considered with care. Nevertheless,
the measured gel points for $N\ge15$ at $r=1$ and $\phi=0.3$ are
delayed by about 5-6 percent of conversion with respect to the mean
field prediction of $0.707$. However, the amount of intra-molecular
reactions of these samples is only about 3\% and thus, not sufficient
to explain quantitatively the observed shift of the gel point.

Lang et al. \cite{Lang2005a} perform Monte-Carlo simulations of network
formation and analyzed the loop size distribution. Here, exponents
$2.5\pm0.1$ and $2.35\pm0.1$ were obtained in the vicinity of the
presumed gel point for two series of data at different degree of dilution
respectively. These exponents are in the same range as the data of
Wang et al. \cite{Wang2017} and more in line with mean field models
as with percolation. In a later publication \cite{Lang2007} by the
same authors, it was recognized that intra-molecular reactions alone
do not explain the shift between data and mean field theory in the
vicinity of the gel point, see Figure 12 and 13 of Ref. \cite{Lang2007}
and the corresponding discussion.

\providecommand{\latin}[1]{#1}
\providecommand*\mcitethebibliography{\thebibliography}
\csname @ifundefined\endcsname{endmcitethebibliography}
  {\let\endmcitethebibliography\endthebibliography}{}

\newpage

\includegraphics[width=1\columnwidth]{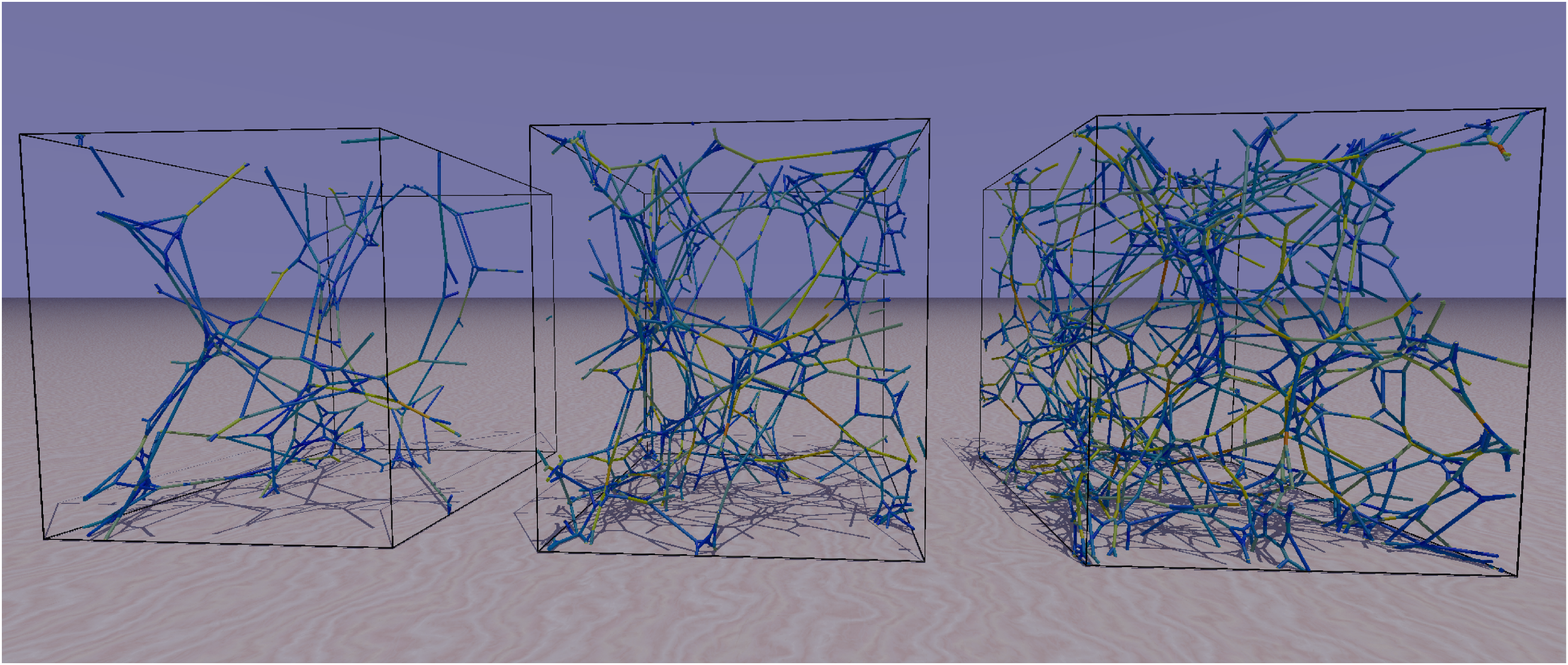}

Table of Contents Graphics: The force balance condition for the determination
of the phantom modulus of a network with junction functionality $f=4$
and chain degree of polymerization $N=8$ is shown at three different
conversions $p=0.66$, $0.67$, and $0.68$ not far beyond the gel
point.
\end{document}